\documentclass{aa}
\usepackage{times}
\usepackage[T1]{fontenc}
\usepackage[latin1]{inputenc}
\usepackage{amsmath}
\usepackage{graphics}	
\usepackage{amssymb}
\usepackage{subfigure}
\usepackage{portland,lscape}	
\usepackage{aabib99,psfig}

\def\2200{2175~\AA~}
\def\Lya{Ly${\alpha}$}

\def\etal{et al. }

\AtBeginDocument{
  
}

\begin{document}

\thesaurus{03 (11.05.2; 11.01.1; 11.01.2; 11.19.3; 02.16.2; 02.19.2)}

\title{Radio galaxies at $z\sim2.5$: results from Keck spectropolarimetry}

\author{Jo\"{e}l Vernet \inst{1}  \and 
Robert A. E. Fosbury \inst{2}  \and 
Montserrat Villar-Mart\'{\i}n \inst{3}  \and 
Marshall H. Cohen \inst{4}  \and 
Andrea Cimatti \inst{5}  \and 
Sperello di Serego Alighieri \inst{5} \and
Robert W. Goodrich \inst{6}}

\mail{J. Vernet. email: jVernet@eso.org}

\institute{European Southern Observatory, Karl Schwarzschild Str. 2, D-85748, Garching-bei-M\"unchen,
Germany \and  Space Telescope European Coordination Facility, Karl Schwarzschild
Str. 2, D-85748, Garching-bei-M\"unchen, Germany \and Dept. of Natural Sciences,
Univ. of Hertfordshire, College Lane, Hatfield, Herts AL10 9AB, UK \and California
Institute of Technology, Mail Stop 105-24, Pasadena, CA 91125, USA \and Osservatorio
Astrofisico di Arcetri, Largo E. Fermi 5, I-50125, Firenze, Italy \and W. M. Keck Observatory 
65-1120 Mamalahoa Highway, Kamuela, HI 96742, USA}

\date{Accepted: 27/10/2000}

\maketitle
\authorrunning{J. Vernet \textit{et al.}}

\begin{abstract}
In classifying the ensemble of powerful extragalactic radio sources,
considerable evidence has accumulated that radio galaxies and quasars
are orientation-dependent manifestations of the same parent population:
massive spheroidal galaxies containing correspondingly massive black
holes. One of the key factors in establishing this unification has been
the signature of a hidden quasar detected in some radio galaxies in
polarized light. The obscuration of our direct view of the active
nucleus usually, but not necessarily exclusively, by a thick nuclear
disk or torus can act conveniently as a `natural coronograph' that
allows a much clearer view of the host of a radio galaxy than of a
quasar.

In this study, we exploit the opportunity to eliminate the quasar
glare by performing sensitive spectropolarimetry with the Keck~II
telescope of a sample of radio galaxies with redshifts around
2.5. This represents the epoch when quasars were many times more
common that they are now and is likely to be the period during which
their host galaxies were being assembled into what become the most
massive galaxies in the Universe today. We show that dust-reflected
quasar light generally dominates the restframe ultraviolet
continuum of these sources and that a highly clumped scattering medium
results in almost grey scattering of the active galactic nucleus 
photons. The observations, however, do not exclude a substantial star
formation rate averaged over a Gyr of evolution. The sub-mm
reradiation from the scattering dust is likely to represent only a
small fraction ($\sim10\%$) of the total far infrared 
luminosity. An analysis of the emission lines excited in the
interstellar medium of the host galaxy by the hard quasar
radiation field reveals evidence of a dramatic chemical evolution
within the spheroid during this epoch. Secondary nitrogen production
in intermediate mass stars produces a characteristic signature in the
N{\sc v}/C{\sc iv} and N{\sc v}/He{\sc ii} line ratios which has been
seen previously in the broad line region of quasars at similar redshifts. We find
intriguing correlations between the strengths of the \Lya~ and N{\sc
v} emission lines and the degree of ultraviolet continuum polarization which
may represent the dispersal of dust associated with the chemical
enrichment of the spheroid.

\end{abstract}

\section{Introduction}

Radio galaxies are thought to be the hosts of radio quasars, oriented
such that the direct view to their nuclei is obscured by an optically
very thick nuclear structure (Antonucci 1993\nocite{antonucci93} and
references therein). This fortuitous occultation of the active 
galactic nucleus (AGN), which would
otherwise outshine an \( L_{*} \) galaxy by a factor of up to a thousand
or so, acts as a convenient 'natural coronograph', allowing the detailed
study of many properties of the surrounding stars and interstellar medium (ISM). 
Our current belief is that the hosts of powerful radio sources in the distant
Universe are destined to become the giant ellipticals of today: the most
massive galactic systems we know (McLure et al. 1999\nocite{mclure99}).
While some may have commenced their formation at very high redshift, it
is clear that the process of assembly is very active at \( z\sim 2.5 \)
(Pentericci et al. 1999, 2000\nocite{pentericci99}
\nocite{pentericci2000}; van Breugel et al. 1998\nocite{breugel98}).
This corresponds to the epoch when luminous quasars appear to have had
their maximum space density (see eg. Pei 1995\nocite{pei95}; Shaver et
al. 1996\nocite{shaver96}; Fan et al. 2000\nocite{fan2000}). Indeed, a
major goal of our programme is to understand any causal relationship
between the formation of the massive black hole and that of the galaxy
within which it resides.

Although we may not see it directly, the presence of the huge luminosity
radiated by the quasar makes itself felt in many ways in the surrounding
galaxy and understanding those processes is essential if we are to deduce the
basic properties and evolutionary state of the host. The radiation field
ionizes a substantial fraction of the gaseous interstellar medium (ISM),
producing emission line halos extending sometimes hundreds of kpc (van Ojik et
al. 1997a\nocite{ojik97a}) and both destroys and accelerates dust (de Young 1998\nocite{deyoung98}). The
ultraviolet (UV) and optical components of the quasar continuum and broad emission
lines can now, thanks to large telescope polarimetry, be seen scattered from
dust and/or electrons far from the nucleus (see eg. di Serego Alighieri et
al. 1996a\nocite{alighieri96a}; Dey et al. 1996\nocite{dey96}; 
Knopp and Chambers 1997b\nocite{knopp97b}; Cimatti et
al. 1998b\nocite{cimatti98b}). Both of these reprocessing mechanisms,
fluorescence and scattering, contribute to the alignment between the
observed optical and radio structures seen in radio galaxies beyond a
redshift of 0.7 or so (Chambers et al. 1987\nocite{chambers87};
McCarthy et al. 1987\nocite{mccarthy87}). Although at optical
wavelengths the effects of scattering are seen most clearly at higher
redshifts, where the dilution by starlight below the 4000~\AA~ break is
small, the process can be seen to operate in low redshift radio
galaxies when they are observed with large telescopes (Cohen et
al. 1999\nocite{cohen99}).

Above 4000~\AA~ in the restframe, the evolved stellar population
contributes substantially to the observed continuum and produces an
infrared (K-band) Hubble diagram with a remarkably small dispersion
(McCarthy 1993\nocite{mccarthy93}, but see references for the higher
redshift sources, eg. Eales et al. 1997\nocite{eales97}; 
van Breugel et al. 1998\nocite{breugel98}). Below 4000~\AA, 
we might expect to see directly a
young stellar population in a starburst arising as a precursor to the
AGN or, alternatively, as a result of its influence on the surrounding
medium. Objects with a redshift around \( z\sim 2.5 \) give optical
spectrographs access to the restframe spectral band which includes the
strong UV emission lines from \Lya~ through C\textsc{iii}{]} and
beyond and also to that part of the continuum which displays
detectable spectral features from OB stars as well as absorption lines
from the ISM. In addition, the spectrum covers the region where
Galactic-type dust displays the 2200~\AA~ signature and the steep
extinction increase below 1600~\AA.

Our programme of Keck observations of \( z\sim 2.5 \) radio
galaxies is designed to obtain high quality spectrophotometry of the
spectral features mentioned above together with polarimetry having
sufficient signal-to-noise to detect a continuum polarization as low
as one or two percent. The polarimetry is used to separate the
directly received light from the scattered radiation while the low-noise
line and continuum spectroscopy can be used to study many properties
of the host galaxy, especially the physical and dynamical state and
the chemical composition of the ISM. Data from the first two sources
observed, 4C$+$23.56 and 4C$-$00.54, are presented in Cimatti et al.
\cite*{cimatti98b} (hereafter paper I). Here, we report the extension of
the sample to nine sources. In this paper, the analysis is restricted to the integrated 
properties defined by the continuum extent. 
An analysis of the spatial variations will come in a later
publication (Villar-Mart\'{\i}n et al. in prep.). Throughout this paper we assume 
$H_{0}=50$ km~s$^{-1}$~Mpc$^{-1}$ and \( q_{0}=0.1 \).

\section{Observations and data reduction}

\begin{table}
\centering 
\begin{tabular}{cccccc}
\hline 
\multicolumn{2}{c}{Object}&
 z&
 Exp. (s)&
 P.A.(\degr)&
 Run\\
 IAU&
 4C&
&
&
&
\\
 (1)&
 (2)&
 (3)&
 (4)&
 (5)&
 (6)\\
\hline 
1243$+$036&
 4C$+$03.24&
 3.560&
 17200&
 156&
 c\\
 0943\( - \)242&
 -&
 2.922&
 13800&
 73&
 b\\
 0828$+$193&
 -&
 2.572&
 18000&
 44&
 b\\
2105$+$233&
 4C$+$23.56&
 2.479&
 22266&
 47&
 a\\
 0731$+$438&
 4C$+$43.15&
 2.429&
 22720&
 12&
 b\\
1410\( - \)001&
 4C\( - \)00.54&
 2.360&
 16240&
 134&
 a\\
1931$+$480&
 4C$+$48.48&
 2.343&
 12000&
 50&
 c\\
 0211\( - \)122&
 -&
 2.340&
 28580&
 104&
 b\\
1809$+$407&
 4C$+$40.36&
 2.265&
 10742&
 82&
 c \\
\hline 
\end{tabular}

\caption{\label{journal}Journal of observations. (1) IAU object name; (2) 4C name (3)
Redshift; (4) Total exposure time; (5) slit position angle; (6) Run a: July
05--06 1997 (data published in Cimatti et al. (1998)) , b: Dec. 24--25--26 1997,
c: May 26--27 1998}
\end{table}

The list of objects and the journal of observations are given in table
\ref{journal}. All sources were selected from the ultra-steep spectrum
(USS) radio galaxy survey (see eg. R\"{o}ttgering et al.
1995b\nocite{rottgering95b}) with redshift greater than 2.2 in order to
observe the \Lya~ line. While the statistical properties of this subset
cannot be defined with precision, it is effectively unbiased with
respect to both radio flux and optical magnitude for the identified
sources (Carilli et al. 1997). The R magnitudes range from 21 to 23 (see
table \ref{results}, column 8). The source 4C$+$03.24, which has a
redshift of 3.6, was originally included as an attempt to investigate
the dependence of properties on epoch but, due to the faintness of many
of the more distant objects, it was subsequently decided to restrict the
sample to $z \leq 3$.

Observations were made using the Low Resolution Imaging Spectrometer (LRIS, Oke
et al. 1995\nocite{oke95}) with its polarimeter (Goodrich et al. 1995b\nocite{goodrich95b})
at the Keck~II 10m telescope from July 1997 to  May 1998 
under subarcsecond seeing conditions (seeing ranging from 0.5\arcsec~to
1\arcsec). The LRIS detector is a Tek 2048\( ^{2} \) CCD with 24\( \mu m \)
pixels which correspond to a scale of 0.214\arcsec pixel\( ^{-1} \). We used
a 300 line mm\( ^{-1} \) grating and 1\arcsec~ wide slit which provide a dispersion
of 2.4~\AA~ pixel\( ^{-1} \) and an effective resolution of \( \sim  \)10~\AA~FWHM.
The spectral range is \( \lambda _{obs}\sim3900-9000\)~\AA. Observations of
each object were divided into several sets (between 2 and 5) of 4 exposures
of approximately 30 minutes with half-wave plate position angle set successively
to 0\degr, 45\degr, 22.5\degr~ and 67.5\degr. The slit was always oriented along
the radio axis (Röttgering et al. 1994\nocite{rottgering94}, Carilli et al. 1997a\nocite{carilli97a}).

After debiasing and flatfielding, 2D-spectra were rectified for geometrical
distortion and wavelength calibrated using Hg-Kr arc lamp spectra. Spectra were
cleaned for cosmic rays hits and then extracted in apertures given in table
\ref{results} in order to include all the extended continuum flux. After performing
a precise wavelength registration of all eight spectra in each set (ordinary
and extraordinary rays for each half-wave plate orientation) using  night-sky emission lines
to determine final wavelength zero-points, data were binned
and then combined as described in Cohen et al. (1995, appendix A\nocite{cohen95}) to
form I, Q and U Stokes parameters. Unbiased values for the fractional polarization
were estimated with the best estimator given by Simmons and Stewart \cite*{simmons85}.
Statistical confidence intervals on fractional polarization and polarization
angle were determined using a Monte-Carlo method taking into account the background
polarization and the detector noise (details of the method in Vernet 2000\nocite{vernet2000}).
Polarized (VICyg12, BD+332642) and unpolarized (HZ44, BD+284211)
standard stars were observed in order to check and calibrate the polarimeter.
Flux calibration was done with the spectrophotometric standard stars: GD248, G191-B2B, Feige~22,
Feige~66, Feige~67. Since several nights were not photometric, the spectra
have been scaled to published HST magnitudes when available (Pentericci et al. 1999
\nocite{pentericci99}, see column 8 in table \ref{results}) without any aperture correction. 
Spectra were corrected for Galactic extinction using Burstein et al.\cite*{burstein82}
maps and adopting the extinction curve of Cardelli \etal \cite*{cardelli89}.

A composite spectrum of all nine  radio galaxies was also
computed. The result is shown in fig. \ref{vary_tau}.  We first normalized the
spectra in a line free region between 2000 and 2100~\AA~ and then
combined them using a simple average.

\section{Results}

\begin{figure*}
{\par\centering \subfigure{\resizebox*{0.9\textwidth}{0.45\textheight}{\rotatebox{90}{\includegraphics{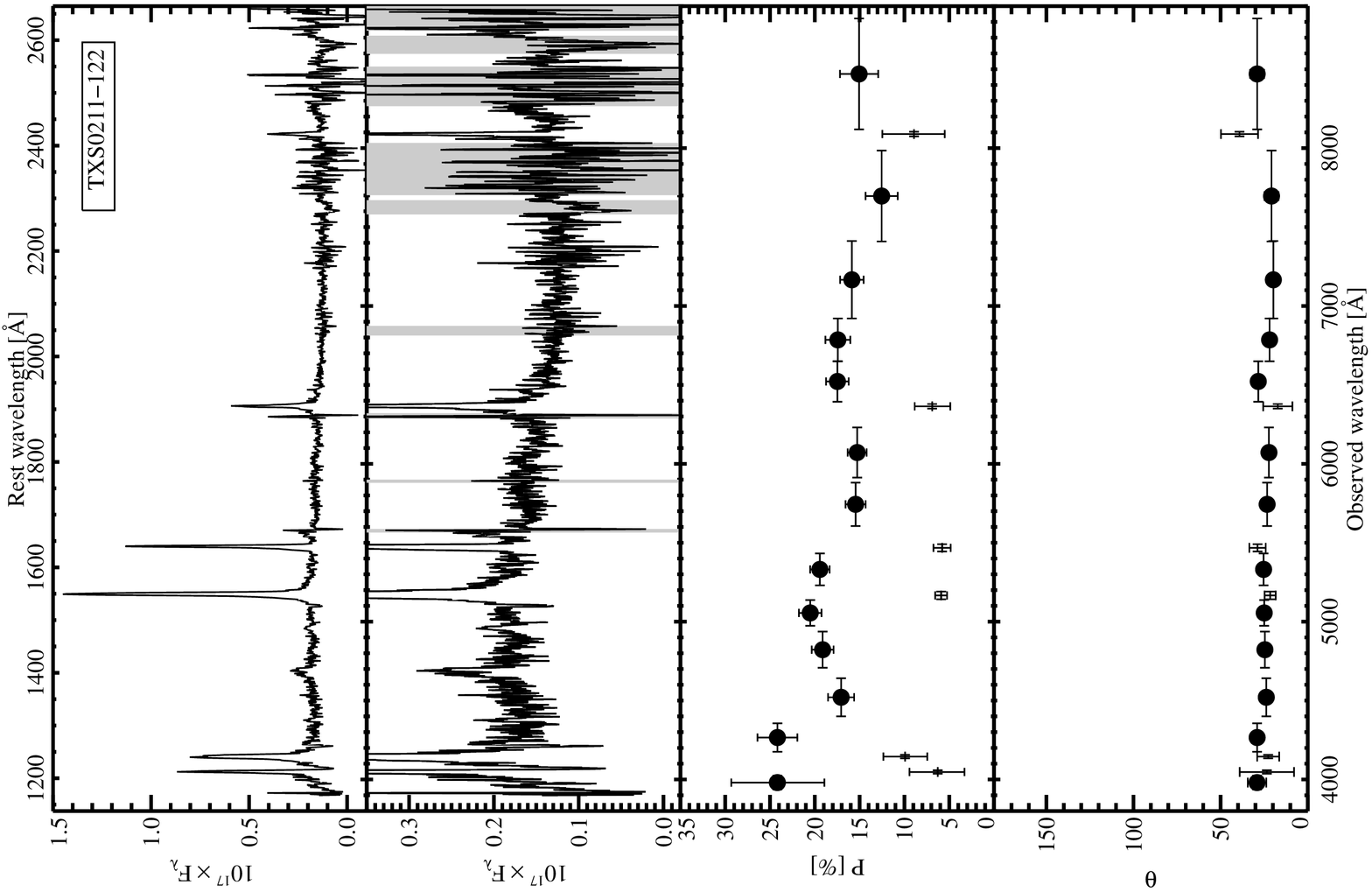}}}} 
\subfigure{\resizebox*{0.9\textwidth}{0.45\textheight}{\rotatebox{90}{\includegraphics{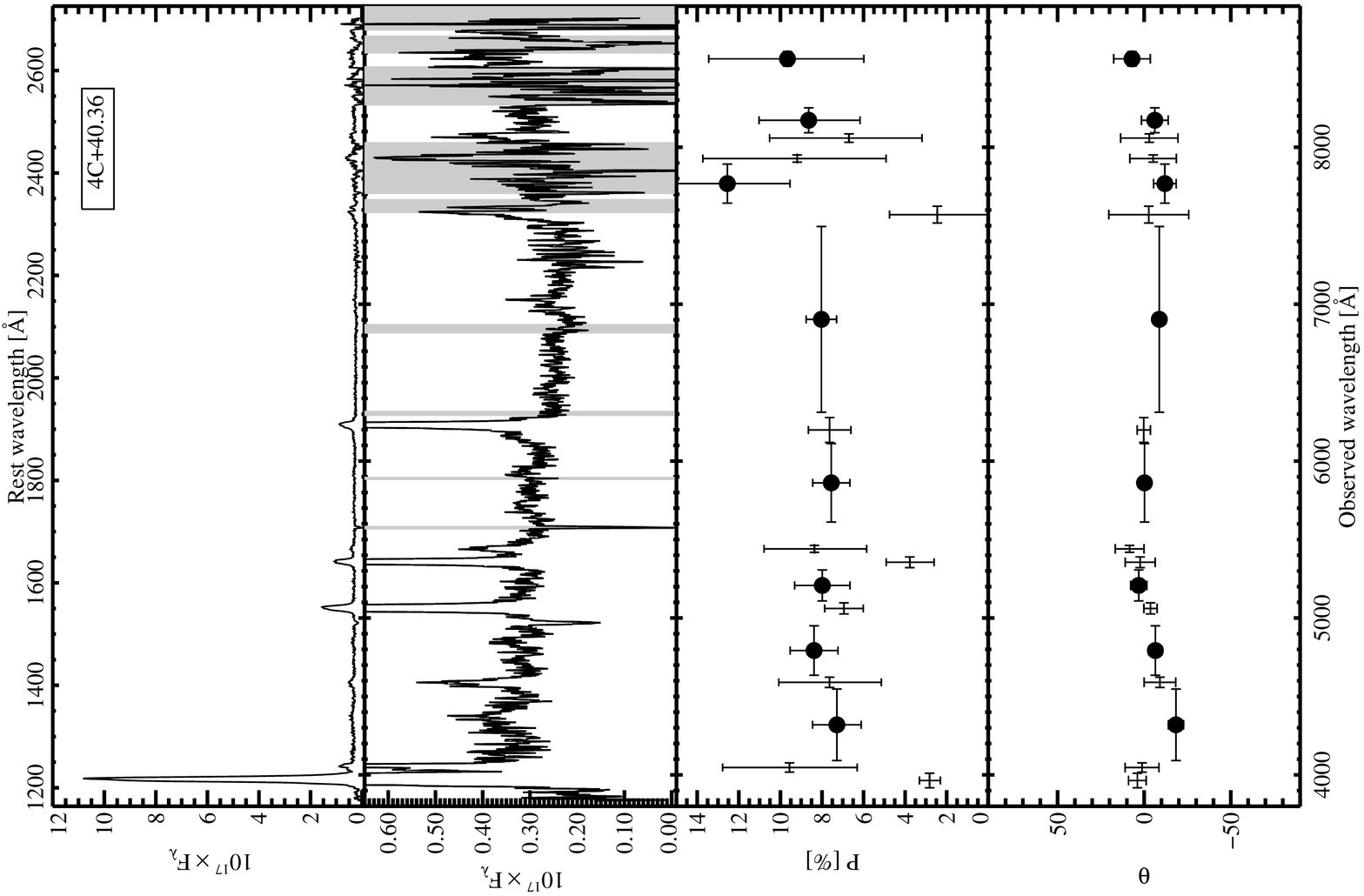}}}} \par}

\caption{\label{polfig} Spectral and polarization properties. In each panel, \emph{from
top to bottom}: the observed total flux spectrum in units of $10^{-17}$erg~s$^{-1}$~cm$^{-2}$~\AA$^{-1}$ plotted at two different scales,
the first to show strong emission lines and the second to show the continuum,
the fractional polarization in \%\ and the position angle of the electric vector (measured N through E). Filled
circles and crosses respectively indicate continuum and narrow emission lines
(including their underlying continuum). \textit{\emph{Shaded areas indicate
regions of strong sky emission.}} \textit{Left : }4C$+$40.36; \textit{Right : }0211$-$122 }
\end{figure*}
 \addtocounter{figure}{-1}
\begin{figure*}
{\par\centering \subfigure{\resizebox*{0.9\textwidth}{0.45\textheight}{\rotatebox{90}{\includegraphics{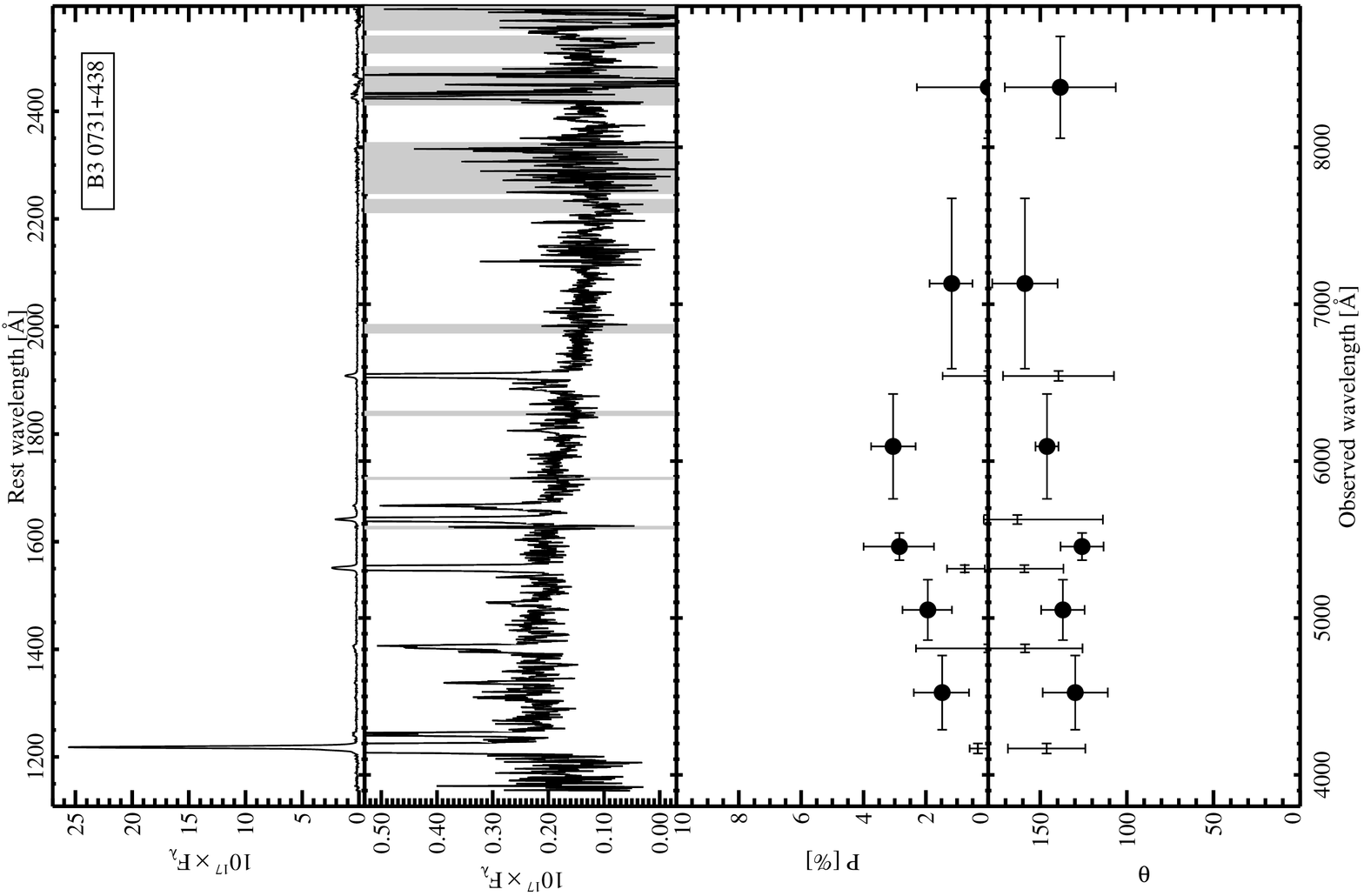}}}} 
\resizebox*{0.9\textwidth}{0.45\textheight}{\rotatebox{90}{\includegraphics{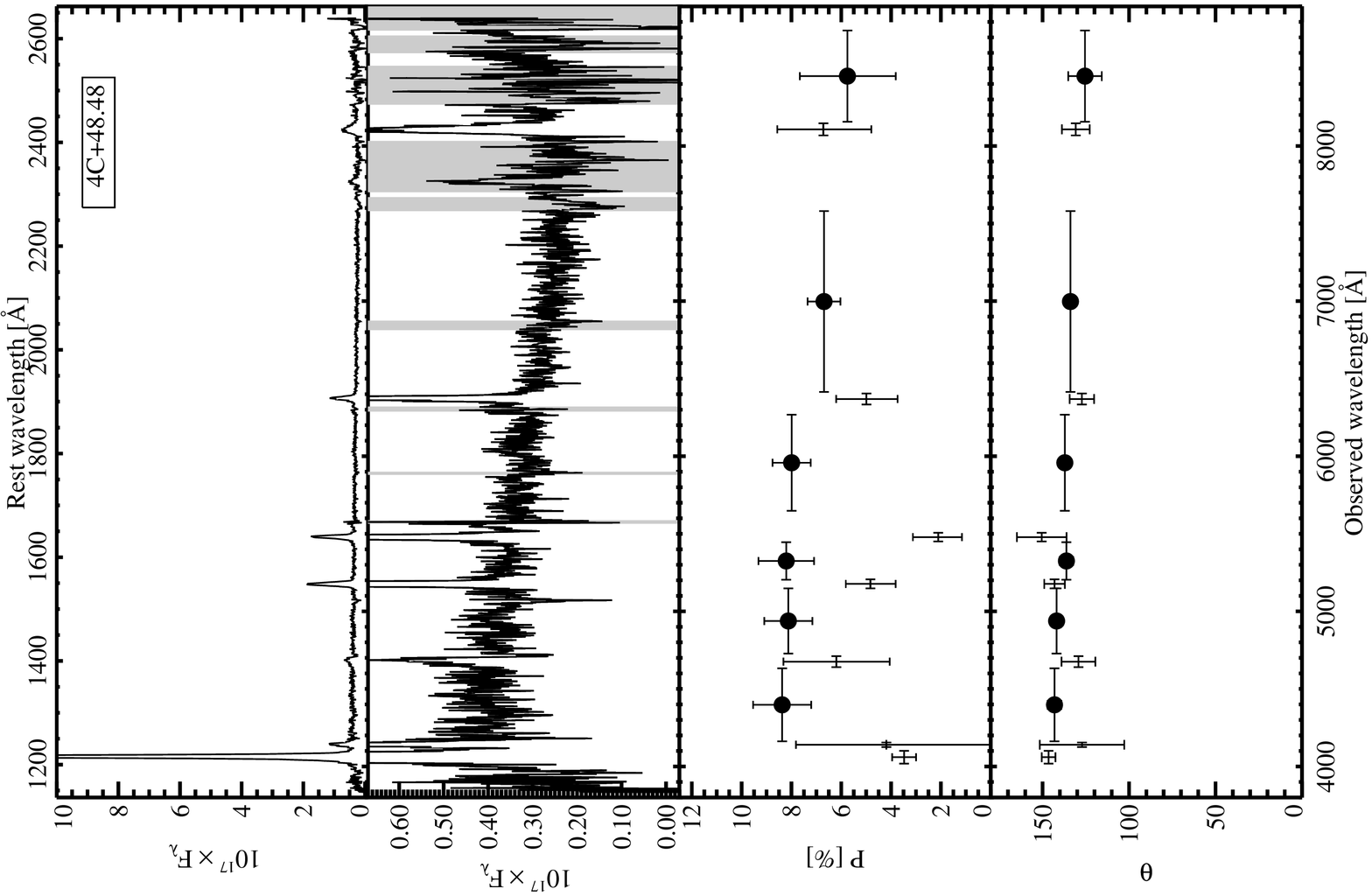}}} \par}

\caption{\textit{Continued. Left : }4C$+$48.48; \textit{Right : } 0731$+$438 }
\end{figure*}
 \addtocounter{figure}{-1}
\begin{figure*}
{\par\centering \subfigure{\resizebox*{0.9\textwidth}{0.45\textheight}{\rotatebox{90}{\includegraphics{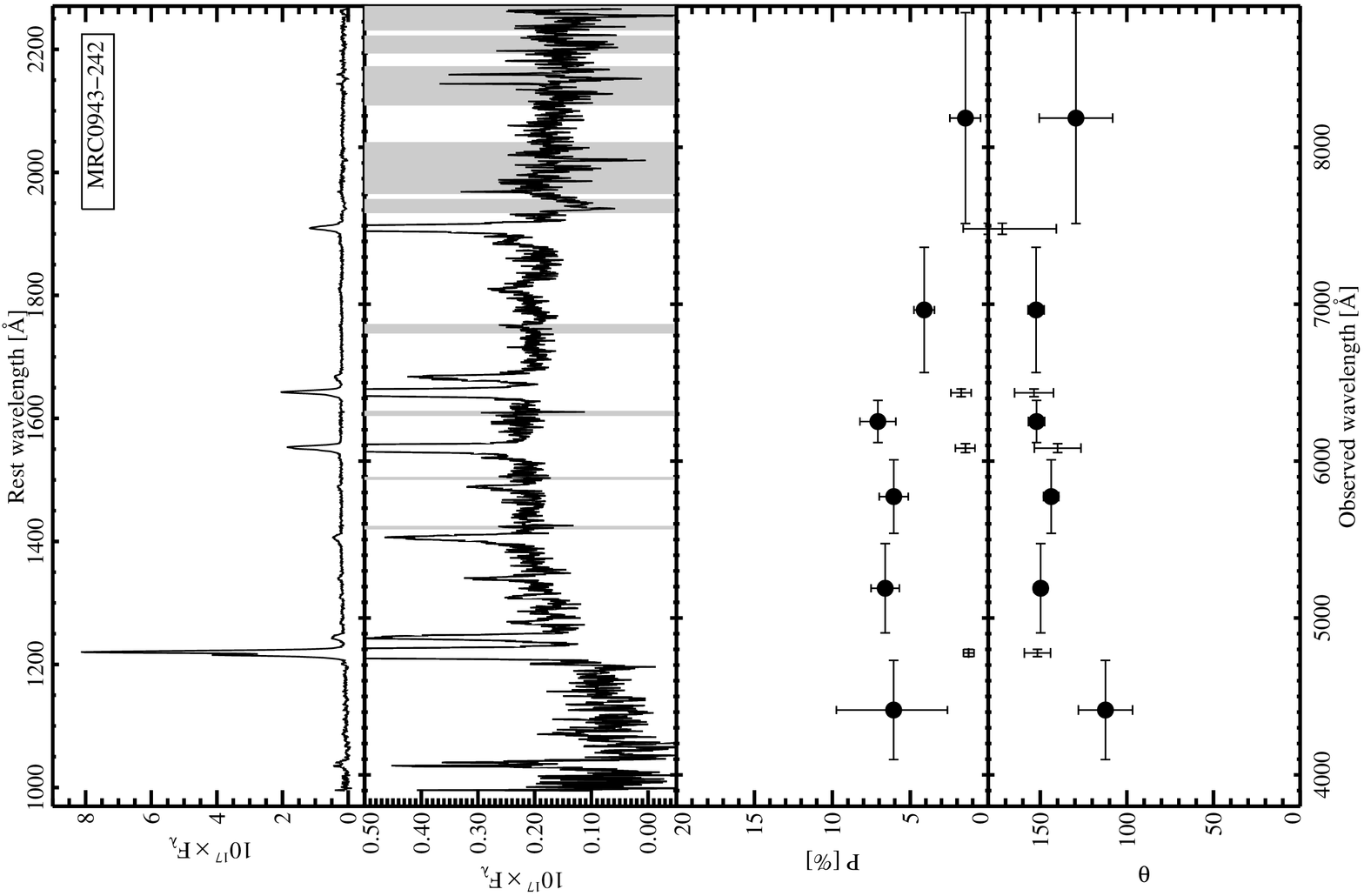}}}} 
\subfigure{\resizebox*{0.9\textwidth}{0.45\textheight}{\rotatebox{90}{\includegraphics{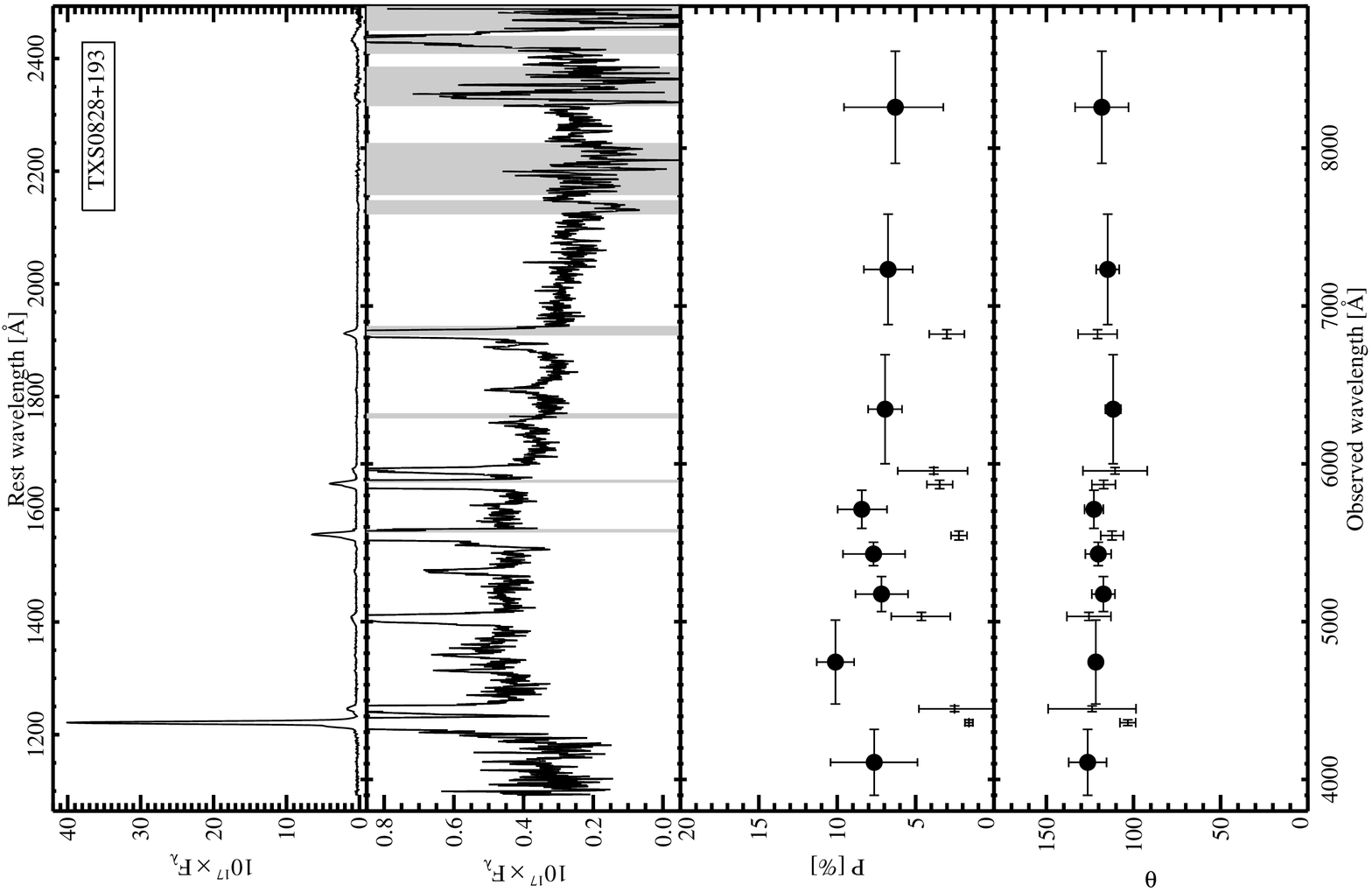}}}} \par}

\caption{\textit{Continued. Left : }0828$+$193; \textit{Right : } 0943$-$242 }
\end{figure*}
 \addtocounter{figure}{-1}
\begin{figure*}
{\par\centering \resizebox*{0.9\textwidth}{0.45\textheight}{\rotatebox{90}{\includegraphics{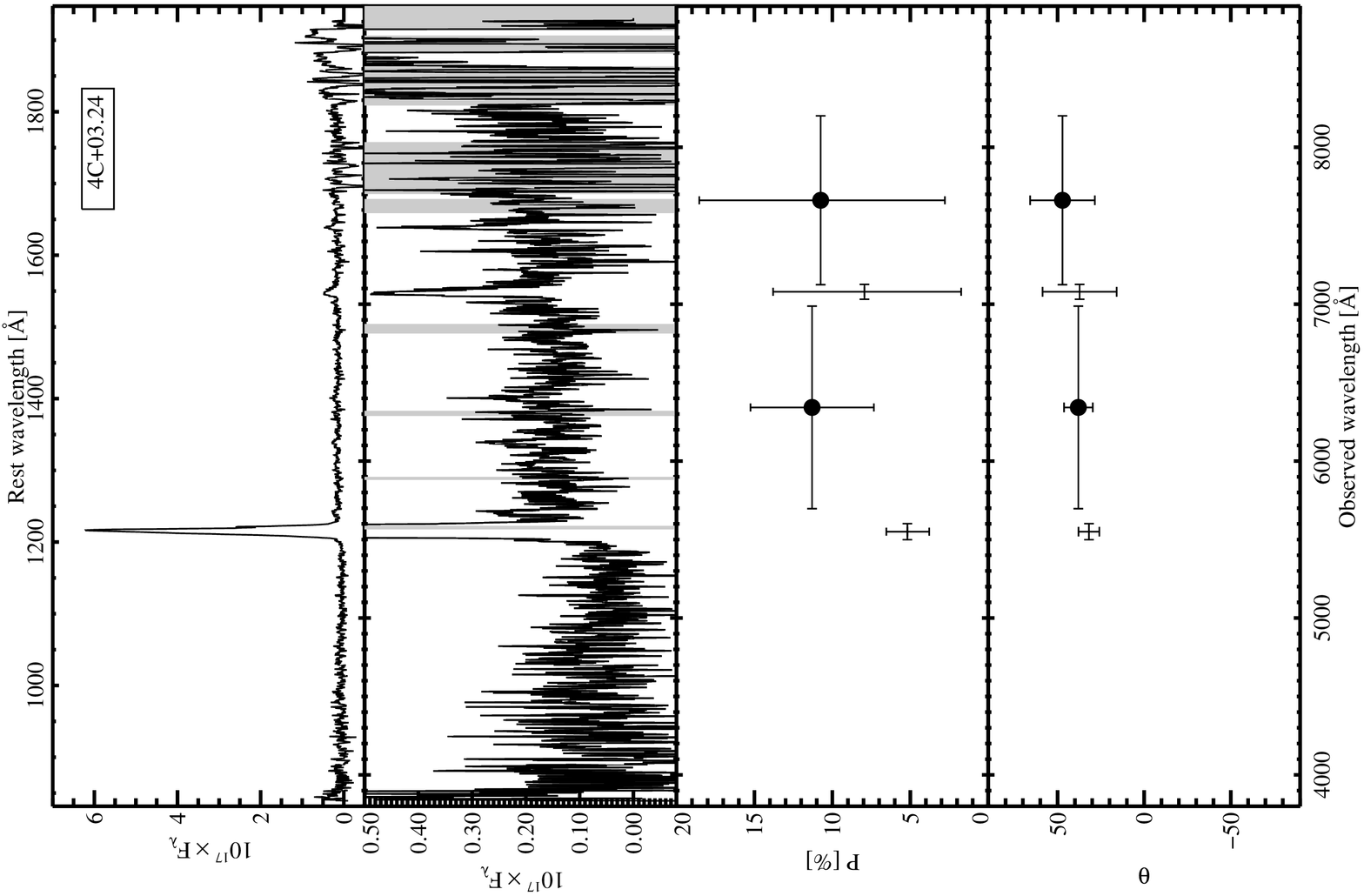}}} \par}

\caption{\textit{Continued.}  4C$+$03.24}
\end{figure*}

\subsection{The continuum}

The spectropolarimetry results are presented in fig.
\ref{polfig}. On each panel is displayed, from top to bottom: the
total flux spectrum at two different scales, the fractional polarization and the position
angle of the electric vector.  The main results are summarized in
table \ref{results}. The aperture used for the extraction matches the
continuum spatial extent (table
\ref{results}, column 2). The continuum is clearly resolved in all
objects extending from 4\arcsec~ up to 8\arcsec~ which corresponds to
a linear scale at $z\sim2.5$ of 45 to 90 kpc. It shows a variety of
morphologies such as double narrow sources (4C$+$23.56), very extended
emission (4C$+$48.48) and single peaked sources (0943$-$242). This
paper does not discuss further the spatial structures but concentrates
solely on the properties integrated over the regions where the
continuum is detected.

The first important result is that all high reshift radio galaxies (HzRGs) 
in this sample except
0731$+$438 show high continuum fractional polarization ranging from \(
\sim  \)6\% up to \( \sim  \)20\% 
measured in a large bin just longward of \Lya~   
(see table \ref{results}, column 3). 
This window, between N\textsc{v} and the 
O\textsc{iv}]/Si\textsc{iv} blend ($\sim1250-1400$~\AA~in the restframe), 
contains several weak, low ionization absorption
lines but is the broadest gap between strong emission lines in a region 
of the spectrum where the instrumental sensitivity is high for all sources. For 0211$-$122
and 0943$-$242, we clearly see a rise of the continuum polarization toward the blue. In
0211$-$122, the fractional polarization rises 
from $P=15.3\pm1.1$\% shortward of C{\sc iii}]
to $P=19.3\pm1.2$\% longward of \Lya~
and for 0943$-$242 it rises from $P=4.1\pm0.6$\%  
to $P=6.6\pm0.9$\% berween the same rest wavelengths 
(note here that error bars are symetric because data were binned
in order to reach signal-to-noise $P/ \sigma > 6$).
A similar trend is also seen in 4C$+$23.56
and marginally in 4C$-$00.54 (see paper I). The signal-to-noise of the
polarization measurements is generally not high enough to detect any
significant variation of fractional polarization with wavelength in
other galaxies in our sample. For 0731$+$438, we can only give an upper
limit of about 2.4\% at the $1\sigma$\ level to the fractional polarization
just longward of the \Lya~ line. However, a significant polarization is measured at
longer wavelength ($P=3.1\pm0.7$\% shortward of C{\sc iii}]).

\begin{figure}
{\par\centering
\resizebox*{1\columnwidth}{!}{\rotatebox{0}{\includegraphics{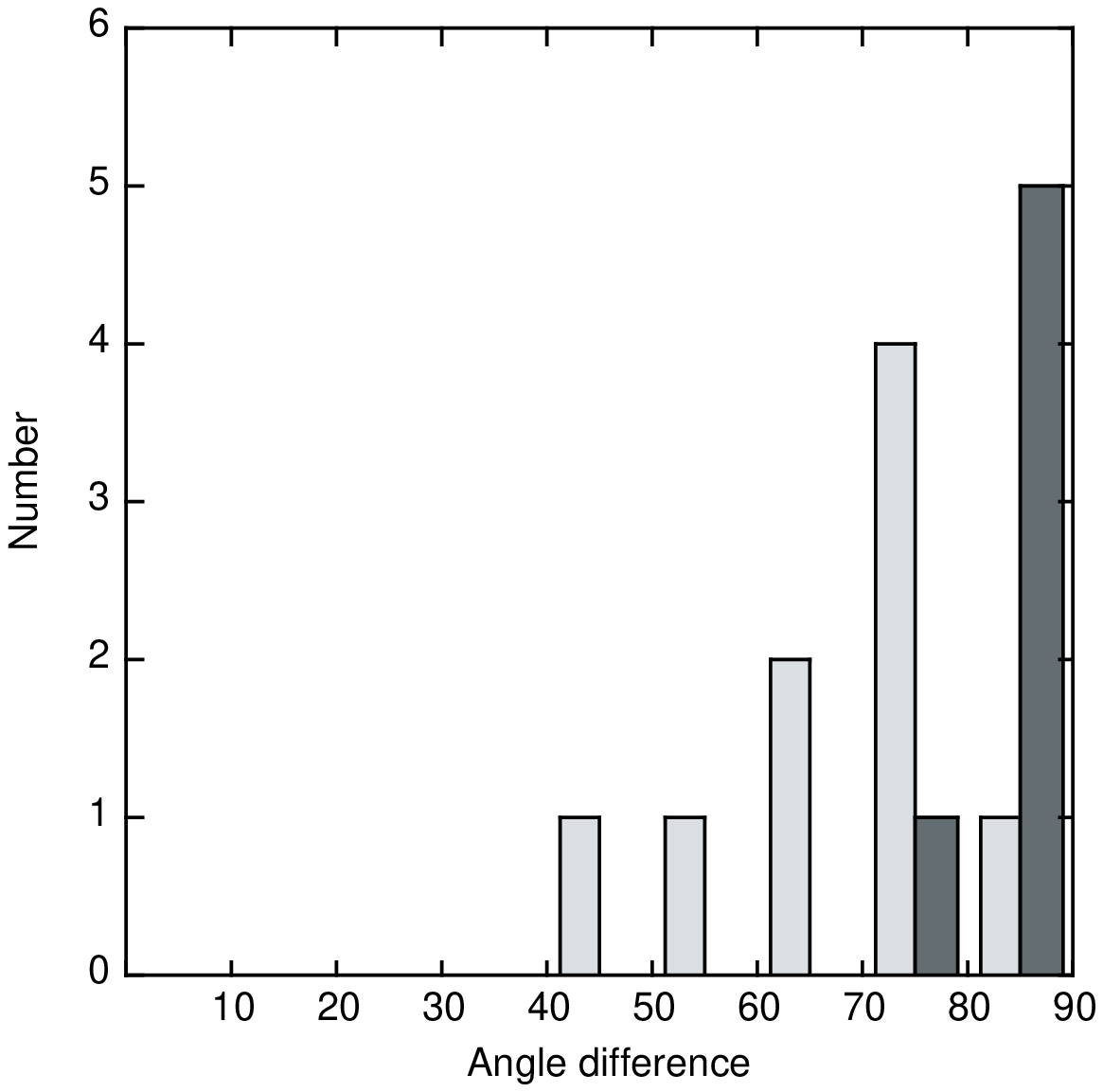}}}
\par}
\caption{\label{padiff} Histograms of the difference between PA$_{rad}$ and PA$_{pol}$ (PA of the E-vector, light) and between PA$_{UV}$ and PA$_{pol}$ (dark).}
\end{figure}

The position angle (PA) of the electric vector (PA$_{pol}$, table \ref{results}, column
4) is generally constant with wavelength within the measurement
errors.  We show in fig. \ref{padiff} the difference between the
radio axis position angle (PA$_{rad}$) and the polarization angle (PA
of the E-vector, light) and the difference between the UV
extension position angle (PA$_{UV}$) and the polarization angle 
(dark). Averaging over the nine sources in our sample we find
PA$_{rad}-$PA$_{pol}\simeq69$\degr. For the 6 sources for which we have
PA$_{UV}$ measurements the average PA$_{UV}-$PA$_{pol}$ is 84\degr.  
This is consistent with  the PA of the
electric vector being generally perpendicular to the UV continuum
extension but less closely correlated to the large-scale radio
structure (this effect was already noted by di Serego Alighieri et al. 1993\nocite{alighieri93}, 
Cimatti et al. 1993,1994\nocite{cimatti93b,cimatti94} 
and Hines \& Wills 1993\nocite{hines93}). 

\begin{table*}
\centering{}
\begin{tabular}{ccccccccc}
\hline 
Object	&Ap.(\arcsec)	&$P$(\%)	&PA\( _{pol} \)(\degr)	&PA\( _{UV} \)(\degr)	&\( \beta  \)	&log(L$_{1500}$)	&R Mag.	&HST Mag. \\
(1)	&(2)	&(3)	&(4)	&(5)	&(6)	&(7)	&(8)&	(9)\\
\hline 
\vspace{0.15cm} 
4C+03.24	&7.1	&11.3$\pm$3.9	&38.0\( \pm  \)8.3	&132		&-	&42.20	&22.5	&23.2 (F702W)\\
 \vspace{0.1cm} 		     
0943\( - \)242	&4.1	&6.6$\pm$0.9	&149.7\( \pm  \)3.9	&67		&-0.84	&42.11	&22.3	&22.6 (F702W)\\
 \vspace{0.1cm} 		     
0828$+$193	&4.1	&10.1$\pm$1.0	&121.6\( \pm  \)3.4	&38		&-1.59	&42.22	&22.7	&22.2 (F675W)\\
 \vspace{0.1cm} 		     
4C+23.56a	&4.9	&15.3$\pm$2.0	&178.6\( \pm  \)3.6	&90		&-0.94	&41.82	&23.2		&\\
 \vspace{0.1cm} 		     
0731$+$438	&7.7	&1.48 ($<$2.4)	&129.8\( \pm  \)19	&-		&-1.43	&41.80	&23.0		&\\
 \vspace{0.1cm} 		     
4C\( - \)00.54	&4.1	&11.7$\pm$2.7	&83.6\( \pm  \)4.1	&6		&-0.88	&41.75	&22.6	&22.9 (F606W)\\
 \vspace{0.1cm}			     
4C+48.48	&6.2	&8.37$\pm$1.5	&143.1\( \pm  \)3.9	&-		&-0.74	&41.96	&22.1 		&\\
 \vspace{0.1cm} 		     
0211\( - \)122	&4.1	&19.3$\pm$1.15	&25.0\( \pm  \)1.8	&122		&-0.93	&41.68	&22.7	&22.9 (F606W)\\
 \vspace{0.1cm}			     
4C+40.36	&4.7	&7.30$\pm$1.2	&161.6\( \pm  \)4.4	&- 		&-0.89	&41.86	&22.4 		&\\
\hline 
\end{tabular}
\par {}

\caption{\label{results} Continuum properties. 
(1): Object name; 
(2): Aperture used for spectral extraction (the slit width was 1\arcsec~ except for 4C$+23.56$ and 4C$-$00.54 for which the slit was 1.5\arcsec~ wide) ; 
(3): Fractional polarization computed in a large bin between N{\sc v}  and Si{\sc iv} lines
 ($\sim$ 1250$-$1400~\AA) with 1 $\sigma$ confidence interval (bins are not necessarily identical to the ones shown on fig. \ref{polfig}) ; 
(4): Electric vector position angle with 1 $\sigma$ confidence interval ; 
(5) Position angle of the UV continuum extension as measured on HST images (Pentericci et al. 1999); 
(6) Slope \( \beta  \) between 1500~\AA~ and 2000~\AA~ (with \( F_{\lambda }\propto \lambda ^{\beta } \)) ;
(7) Logarithm of monochromatic luminosity at \( \lambda _{rest}\approx  \)1500~\AA~ in erg~s$^{-1}$\AA$^{-1}$ ; 
(8) R band magnitude from Carilli et al. (1997) and R\"ottgering et al. (1995); 
(9) HST magnitude from Pentericci et al. (1999).
}
\end{table*}

One very striking result is that all objects in this sample show a
remarkably similar continuum shape with a dip around 2200~\AA~ and a
shallow peak in $F_{\lambda}$ between \Lya~ and 1500~\AA. This similarity of shape and in particular
the 2200~\AA~ dip cannot be an artifact of the response calibration
since objects in the sample cover a relatively large range in
redshift. For instance, 0943$-$242 at $z=2.922$
($1000<\lambda_{rest}<2250$) does not show any dip between observed wavelength 7000~\AA~ and 8000~\AA. Although spectra are quite noisy at long wavelength, the
continuum rise at \( \lambda _{rest}>2200 \)\AA~ is well detected
using just regions between strong sky lines (see in between shaded areas showing 
region of strong sky emission in fig. \ref{polfig}). The decrease below
1500~\AA~ is also a real continuum feature since it starts before the
onset of the Lyman forest absorption below \Lya. Continuum slopes
between 1500~\AA~ and 2000~\AA~ are listed in table \ref{results},
column 6.

\subsection{\label{section_emlines}Emission lines}

Fluxes of all detected emission lines are given in table 
\ref{em_lines}.  The quality of the data allow us to detect a number
of weak emission lines such as Si\textsc{ii}\( \lambda \)1309,
N\textsc{iv}]\( \lambda
\)1488 and N\textsc{iii}]\( \lambda  \)1750. The spatial extent of strong
emission line halos (\Lya, N\textsc{v}, C\textsc{iv} and
He\textsc{ii}) ranges from about 45 kpc in 0943$-$242 up to more than
100 kpc in 0211$-$122 and 0828$+$193 (see also van Ojik et al. 1996\nocite{ojik96}). These strong emission lines are usually more
extended than the aperture used for the extraction and show very
complex spatial and velocity structure with high velocity amplitudes up
to \( \sim \)2000 km~s\( ^{-1} \). 
Narrow emission lines have typical width of \( \sim \)800$-$1500 km~s\( ^{-1} \) as
measured on the He{\sc ii}$\lambda$1640 line (see table \ref{heii_fwhm}) which is less affected by
possible underlying broad emission (since it is usually very weak in quasars) or absorption. 
Broad lines are clearly detected in 0211$-$122
(broad C\textsc{iv} FWHM $\sim6500$ km s \( ^{-1} \), $W_{\lambda}\sim57$ \AA~ and broad C\textsc{iii}{]} FWHM $\sim9000$
km s\( ^{-1} \), $W_{\lambda}\sim50$ \AA) and 4C$+$23.56 (broad C\textsc{iv} FWHM $\sim4800$ 
km s\( ^{-1} \), $W_{\lambda}\sim68$ \AA~ and broad C\textsc{iii}{]} FWHM $\sim6300$ km
s\( ^{-1} \), $W_{\lambda}\sim76$ \AA). Broad wings are also marginally detected in 4C$+$40.36
and 4C$-$00.54.

Since narrow emission line profiles can be quite poorly represented by
gaussians due to the presence of several kinematic components, we did
not attempt to perform gaussian fitting. Fluxes were measured from
simple line integrals. The main source of error in these measurements
comes from uncertain continuum determination. We estimated the
uncertainty by making limiting high and low estimates of the continuum
level.  This method does not provide rigorously defined measurement
errors but does give a conservative estimate of the range in flux
allowed by the data for each emission line.

One particular issue in measuring narrow emission line fluxes is the
contamination by broad components, especially for the Ly\( \alpha  \),
N\textsc{v}, C\textsc{iv} and C\textsc{iii}{]} lines. In the two cases
where broad lines are most prominent, (4C+23.56 and 0211$-$122) and most
likely to significantly affect our results, we made measurements of the
narrow emission lines using several different strategies including
multiple-gaussian fitting, integration from an elevated continuum and
measurement after subtraction of a scaled quasar template. The small
dispersion in the resulting values is included in our conservative
estimate of the uncertainties.

We also measured the polarization of the strongest narrow emission lines including
the underlying continuum (see bins
marked with crosses in fig. \ref{polfig})\@. They show, in general, a significantly
lower polarization than neighbouring pure continuum bins. In all cases the continuum
subtracted polarization in the narrow Ly\( \alpha  \) line is less than 2\%. 

\begin{table}[h!]
\centering{
\begin{tabular}{lccc}

\hline
Source	          &$W_{\lambda}$  &\multicolumn{2}{c}{FWHM}\\
                &(\AA)         &(\AA) & ($km\,s^{-1}$)    \\
\hline

4C$+$03.24      &29  $\pm$ 11    &19.6        &784  \\
0943$-$242      &216 $\pm$ 20    &22.4        &1042  \\
0828$+$193      &243 $\pm$ 12    &26.7        &1366 \\
4C$+$23.56a     &69  $\pm$ 9     &14.3        &749 \\
0731+438        &152 $\pm$ 9     &10.5        &560  \\
4C$-$00.54      &109 $\pm$ 9     &14.5        &788 \\
4C$+$48.48      &97  $\pm$ 13    &16.9        &922 \\
0211$-$122      &84  $\pm$ 4     &11.4        &626  \\
4C$+$40.36      &74  $\pm$ 10    &29.6        &1655 \\
\hline

\end{tabular}
}

\caption{\label{heii_fwhm} Equivalent width and instrumental profile (10~\AA) corrected observed FWHM 
measurements of the HeII $\lambda$ 1640 emission line. FWHM measurements were done
using a single gaussian fit in all cases without any attempt to deblend the different kinematic components.}

\end{table}

\subsection{Absorption lines}

\begin{figure}
{\par\centering \resizebox*{1\columnwidth}{!}{\rotatebox{90}{\includegraphics{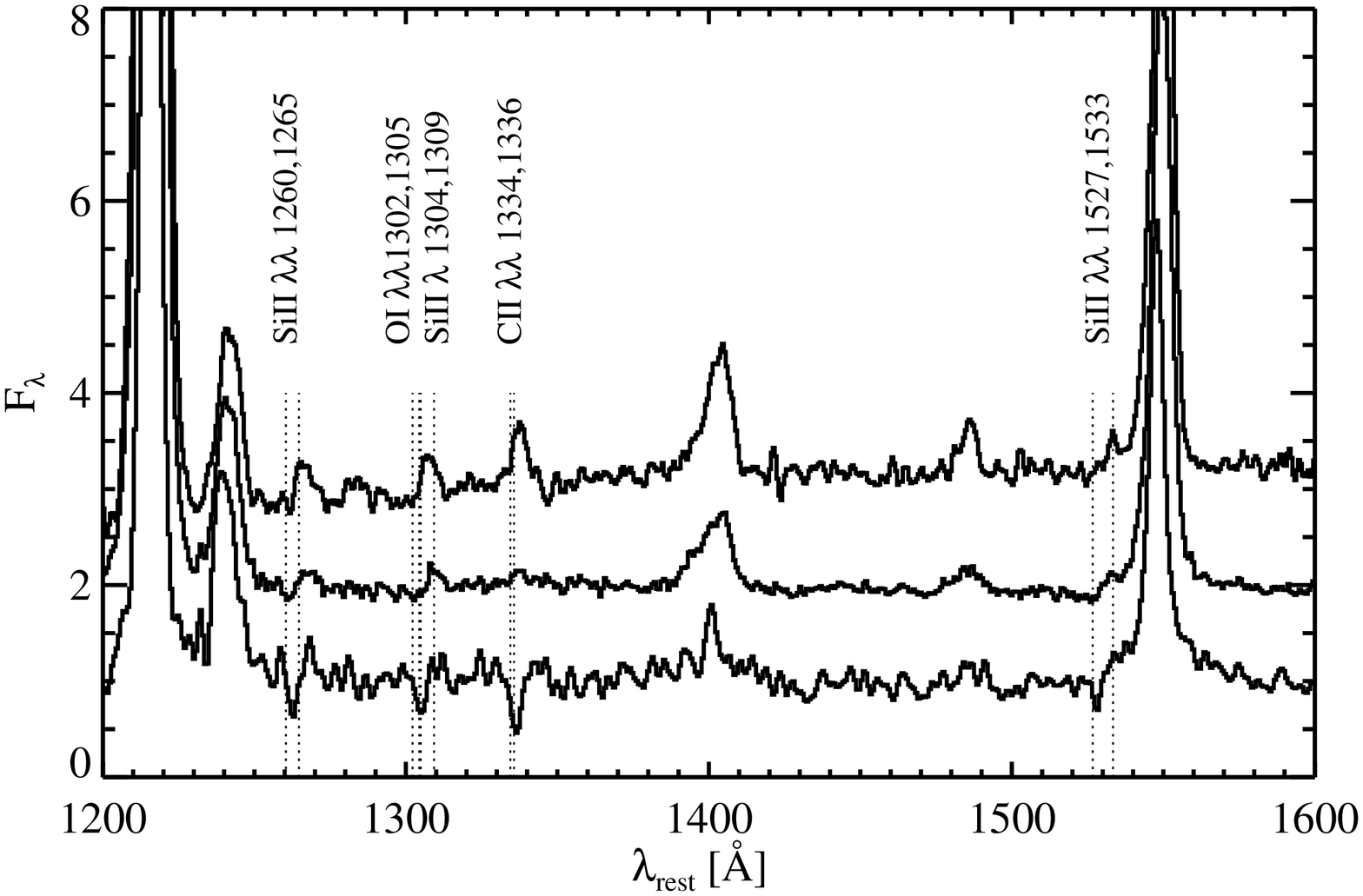}}} \par}
\caption{\label{figabs}Low ionization absorption and emission features for two extreme cases. While low ionization lines are only seen in absorption in 4C$+$23.56 (\textit{bottom spectrum}) these lines only show up in emission in 0943$-$242 (\textit{top spectrum}). The spectrum in the middle is our HzRG composite spectrum.}
\end{figure}

Spectra generally show several absorption features that can be identified as low ionization zero-volt or excited fine-structure interstellar lines such as 
Si\textsc{ii\( \, \lambda\lambda \, \)1260,1265}, 
O\textsc{i}\( \, \lambda\lambda \,  \)1302,1305 
and Si\textsc{ii}\( \, \lambda\lambda \,\) 1304,1309, 
C\textsc{ii}\( \, \lambda\lambda \, \)1334,1336,  
Si\textsc{ii}\( \, \lambda\lambda \, \)1527,1533.  
There is a weak tendency for high
polarization objects to show these lines in emission and low
polarization objects to show them in absorption. 
Equivalent width measurements and identifications of these features are given in table
\ref{low_ion_lines}. Fig. \ref{figabs}
illustrates the complex behavior of these lines with the two extreme cases observed
in our sample. While low ionization lines are only seen in absorption
in 4C$+$23.56 (bottom spectrum in fig. \ref{figabs}), these lines only
show up in emission in 0943$-$242 (top spectrum in fig. \ref{figabs}). 

In the case of 0731$+$438 and 4C$-$00.54 (see paper I), an
absorption is detected close to the expected position of the C\textsc{iii}\( \, \lambda
\lambda \, \)1426,1428 photospheric line but we consider this
identification as very uncertain since no other absorption lines
compatible with pure photospheric lines are detected. See sect. \ref{ysp}
for a more detailed discussion on photospheric absorption lines.

\subsection{Notes on individual objects}

\begin{itemize}
\item 0828$+$193: this radio galaxy was thought to have two subcomponents  separated by 3\arcsec~(OI1 to the SE and OI2 to the NW, see Knopp and Chambers 1997\nocite{knopp97b}).
Our deep spectrum reveals that OI1 is in fact a M5V type star. The 3\( \sigma  \)
upper limit to the polarization of this star between 5600 and 8300~\AA~ is 0.85
\%. 
\item 4C$+$40.36: Chambers et al. (1996b)\nocite{chambers96b} note that there is a small
extension 2\arcsec~ to the southeast of the eastern lobe which is detected in
all bands and has very similar colours to 4C$+$40.36. They identify it as a good
candidate for a companion or a subcomponent (see their fig. 7). This is in
fact a foreground H\textsc{ii} galaxy at \( z=0.404 \). The spectrum is displayed
in fig. \ref{inter_obj}a with an overlay of the H\textsc{ii} galaxy NGC~6052 from
Schmitt et al. (1997)\nocite{schmitt97}. 
\begin{figure*}
{\par\centering \subfigure{\resizebox*{0.49\textwidth}{!}{\rotatebox{90}{\includegraphics{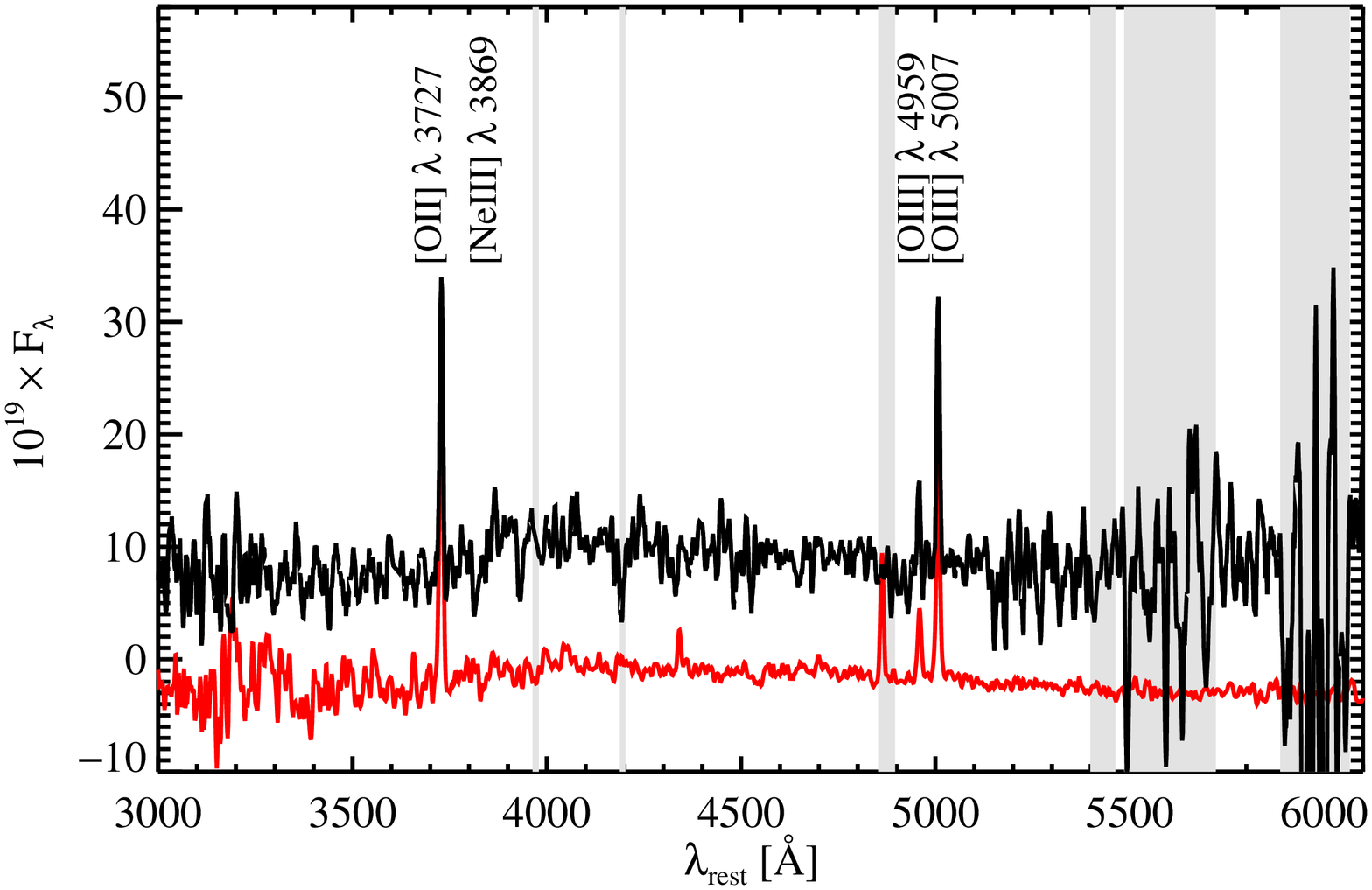}}}} 
\subfigure{\resizebox*{0.49\textwidth}{!}{\rotatebox{90}{\includegraphics{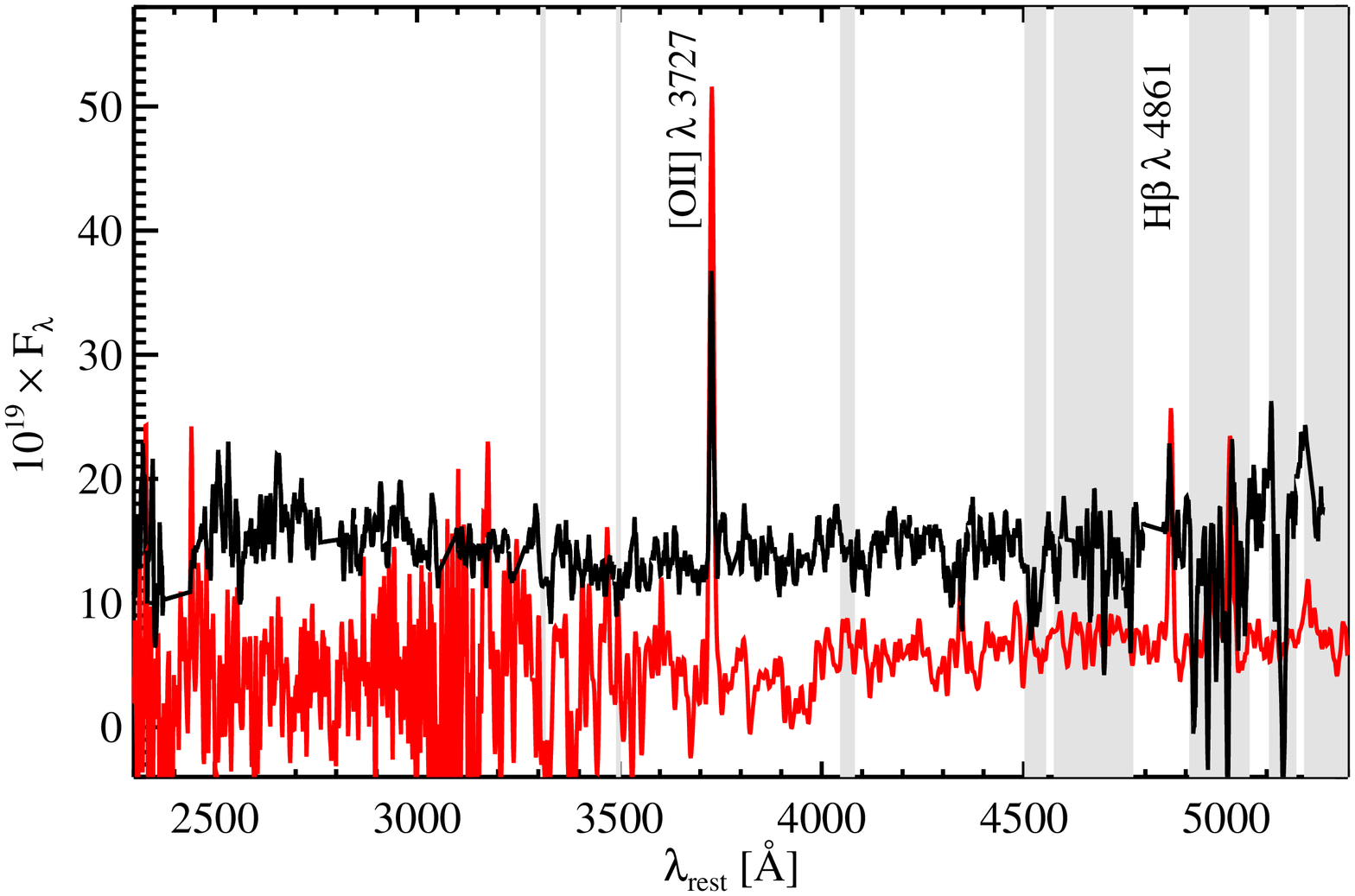}}}} \par}
\caption{\label{inter_obj}Spectrum of intervening galaxies (upper spectrum). 
{\it left} : H\textsc{ii} galaxy 2~\arcsec~ southwest of 4C$+$40.36 with
template galaxy NGC~6052 overlayed; 
{\it right} : H\textsc{ii} galaxy 2.5~\arcsec~ southwest of 4C$+$48.48
with template galaxy NGC~6764 reddened with E$_{B-V}$=0.1
The main emission lines are labeled. In both cases, spectra have been
cleaned of residual extended emission lines from the radio galaxy and
smoothed with a 5 pixel boxcar. Template galaxies were obtained from
Schmitt et al. (1997) and have been shifted downward by 10 flux units
for clarity. Shaded areas show regions of strong sky emission.}
\end{figure*}
\item 4C+48.48: we find an emission line object 2.5~\arcsec~ to the southeast of 4C+48.48
that has a significantly redder continuum than the radio galaxy. The measured
polarization between 5600 and 8000~\AA~ is compatible with 0\% with a 3\( \sigma  \)
upper limit of 4.5\%. A strong emission line is clearly detected at \( \lambda _{obs}= \)
6277.5 \AA. We also identify another weaker emission line at \( \lambda _{obs}= \)
8184.7 \AA. These lines do not correspond to any expected strong emission at
the redshift of the radio galaxy. Identifying these lines as {[}O\textsc{ii}{]}\( \lambda 3727 \)
and H\( _{\beta } \), this intervening object can be classified as an H{\sc ii} galaxy
at $z=0.684$. The spectrum is displayed
in fig. \ref{inter_obj}b with an overlay of the H\textsc{ii} galaxy NGC~6764 from
Schmitt et al. (1997)\nocite{schmitt97}. 
\end{itemize}

Note that, due to the presence of intervening objects along the line of sight,
observed properties of 4C$+$40.36 and 4C$+$48.48 might be significantly affected by lensing.

\subsection{Additional data}

We also collected information on three high redshift sources that have comparable
quality spectropolarimetry data from the literature: two radio galaxies (4C$+$41.17
and 3C~256) and the ultraluminous infrared galaxy FSC~10214$+$4724. The main properties
of these objects that are used in the discussion are given in table \ref{literature-results}.

Six sources in our sample have been observed with HST/WFPC2. These data are
published in Pentericci et al. \cite*{pentericci99}. They were mainly used in
this work to compare the polarization angle to the direction of the extended
ultraviolet emission (see table \ref{results}) and to obtain an absolute flux
calibration of our spectra.

\begin{table*}
\centering{}
\begin{tabular}{ccccccc}
\hline 
Object&
 \( z \)&
$P$(\%)&
\( \beta  \)&
  \Lya/C\textsc{iv}&
 N\textsc{v}/C\textsc{iv}&
 He{\sc ii}/C{\sc iv}\\
\hline 
\vspace{0.15cm} FSC~10214\( + \)4724&
 2.282&
 26\( \pm  \)2&
&
 0.276\( \pm  \)0.003&
 0.736\( \pm  \)0.001&
 0.496\( \pm  \)0.001\\
 4C\( + \)41.17&
 3.798&
 \( \leq  \)2.4 (2\( \sigma  \))&
-1.8&
 11.1\( \pm  \)0.7&
 0.29\( \pm  \)0.03&
 0.42\( \pm  \)0.02\\
 3C~256&
 1.824&
 10.9\( \pm  \)0.9&
-0.71&
 10.4\( \pm  \)0.2&
 0.27\( \pm  \)0.17&
 1.05\( \pm  \)0.02 \\
\hline 
\end{tabular}
\par {}

\caption{\label{literature-results} Polarization and line ratio results from the literature
for three sources which have similar quality spectropolarimetry. The data are
taken from: FSC~10214\protect\( +\protect \)4724, Goodrich et al. (1996);
4C\protect\( +\protect \)41.17, Dey et al. (1997); 3C~256, Dey et al. (1996),
Simpson et al. (1999). The fractional polarization measurement has been chosen
to match as closely as possible the aperture and waveband we have used for our
sample sources. In the case of 3C~256, no polarization measurement has been
made shortward of C\textsc{iv}, but Dey et al. (1996) show that \protect\( P\protect \)\ is
quite flat from 1500--2500~\AA. When forming line ratios, we have excluded components
with a FWHM\protect\( \geq \protect \)3,000~km~s\protect\( ^{-1}\protect \). }
\end{table*}

\nocite{goodrich96}\nocite{dey97a}\nocite{dey96}\nocite{simpson99}

Radio properties of all objects in our sample can be found in Röttgering et
al. \cite*{rottgering94} and Carilli et al. \cite*{carilli97a}.

\section{Discussion}

The discussion is based on the integrated properties of the
sources. The spatial properties and kinematic details are addressed in
a separate paper (Villar-Mart\'{\i}n et al. in prep.\nocite{villar2000}). 
The aim is to identify the origins of the
various radiated continuum and line components using polarimetry as a
powerful discriminent between direct and scattered radiation.

\subsection{The nature of the continuum}

The question we address in this section is the origin of HzRG ultraviolet
continuum: what is the dominant source of UV continuum and what is the relative
contribution of the different possible components? The high fractional polarization
together with the presence of broad lines clearly suggest that scattering of
the radiation from the central active nucleus plays an important rôle, and probably
a dominant one in the most polarized objects. The universality of the shape of the
spectral energy distribution is difficult to reconcile with the dominance
of a young stellar population in this wavelength range since starbursts
show strong colour evolution and a range of reddening. However, there is
evidence, mainly from the less strongly polarized objects, that a young stellar population
can make a significant contribution to the UV continuum.

\subsubsection{\label{sect_scat}The scattered continuum}

The high continuum polarization shows that a large fraction of
the UV continuum must be due to scattered AGN light in all objects in our sample
except possibly for the least polarized source (0731$+$438). This case is strengthened
by the clear detection of broad lines in the most polarized objects (0211$-$122,
4C+23.56) with equivalent widths which are similar to those seen in quasars. 

The remarkable similarity of the shape of the HzRG ultraviolet spectral energy distribution 
(SED) with that of quasars
shows that the scattering process must be approximately independent of wavelength.
The most natural explanation for such `grey' reprocessing is that the dominant
mechanism is Thompson scattering by electrons. However, this interpretation encounters
several difficulties. Scattering by a population of hot (\( T_{e}>10^{7}K \))
electrons is ruled out because the Doppler broadening would smear out the scattered
broad lines. In the two objects in our sample for which we can easily measure
the width of the scattered broad C\textsc{iv\( \lambda 1550 \)} line we can
estimate an upper limit on $T_{e}$\ of $\sim10^{5}K$\ (see Miller, Goodrich and
Mathews 1991\nocite{miller91} and Cimatti et al. 1996\nocite{cimatti96}) assuming a typical
value of \( \sim 6500\,$km$\,$s$^{-1} \) for the width of the C\textsc{iv\( \lambda 1550 \)}
line of a radio loud quasar (as measured on Cristiani and Vio 1990 \nocite{cristiani90}
radio loud quasar composite spectrum). In addition, estimates
of the expected X-ray fluxes for such a population of hot electrons in distant
3CR radio galaxies are in strong disagreement with observations (see for instance
calculations made by Dey et al. 1996\nocite{dey96} for 3C~256). Scattering by a cooler
(\( T_{e}\sim 10^{4}-10^{5}K \)) population of electrons as the dominant mechanism
is more difficult to rule out. The most compelling argument is based on the far greater 
scattering efficiency per unit mass of dust than electrons which implies that, as 
soon as some dust is present in the ISM, it will dominate the scattering process, an assumption we maintain throughout the rest of this discussion.

Standard homogeneous optically thin dust scattering models predict a
bluening of the emergent spectrum due to the wavelength dependence of the scattering
efficiency of small dust grains. In order to reproduce observations,
previous modeling has been based on optically thin scattering followed
by arbitrary extinction by a screen of dust. The implicit assumption is
that some part of the dusty medium is illuminated by the quasar and
scatters light to the observer while another part is in the `shadow' and
produces extinction of the scattered light (see Manzini and di Serego
Alighieri 1996\nocite{manzini96}). A more realistic model employs the same
population of dust grains to scatter and absorb the light whereupon the
emergent scattered light is maximized where a photon experiences
scattering and extinction optical depths close to unity (\( \tau
_{scat}\sim \tau _{ext}\sim 1 \) ). If the medium is inhomogeneous
(clumpy), a natural luminosity weighting process operates by which the
emergent flux over a wide range of wavelength comes from paths with
optical depth close to one. The scattered spectrum is then approximately
the input spectrum multiplied by the dust albedo (the ratio of the
scattering to the extinction efficiency) which, for Galactic dust is
almost grey at ultraviolet and optical wavelengths.

\begin{figure*}
{\par\centering \resizebox*{0.98\textwidth}{!}{\rotatebox{90}{\includegraphics{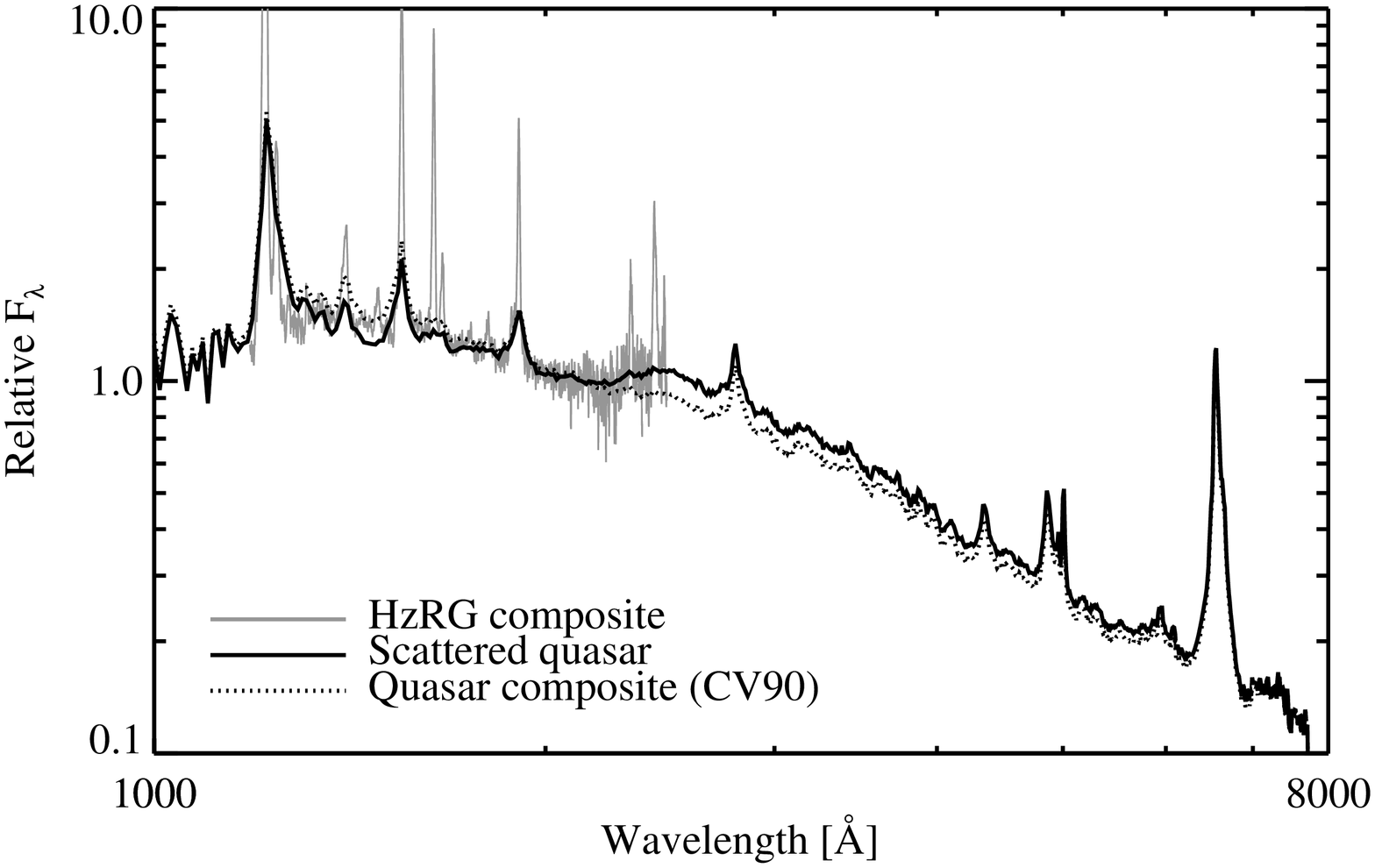}}} \par}

\caption{\label{vary_tau} Comparison of the HzRG composite spectrum (in grey) with
quasar radiation scattered by a clumpy medium (continuous black line) modeled
using V\'arosi and Dwek (1999) models with high density contrast (\protect\( \alpha =1000\protect \))
and small clump volume filling factor \protect\( f_{c}=0.01\protect \). This medium produces
an almost grey scattering illustrated by the similarity between the scattered
spectrum and the input quasar composite spectrum (dotted line). 
The difference between the quasar and the scattered spectrum around 2200~\AA~ is 
due to the dust albedo feature. Note how well the
HzRG composite spectrum tracks this feature. Spectra are normalized at 2000~\AA\emph. }
\end{figure*}

Radiative transfer in a clumpy dusty medium has been extensively studied (see
eg. Witt and Gordon 1996,1999\nocite{witt96,witt99a}; V\'arosi and Dwek 1999\nocite{varosi99}
and references therein) and applied to several types of astrophysical objects
such as circumstellar shells and starforming galaxies. The expected greyness
of the scattering over a large wavelength range described above is confirmed
by all studies. We compared our observations to results obtained using the analytical
model developed and tested against Monte-Carlo simulations by V\'arosi and Dwek
\cite*{varosi99}. These analytical approximations, based on an extension of
the mega-grain model of a two-phase medium first proposed by Hobson and Padman \cite*{hobson93b}
allowed us to easily explore the main parameters describing the clumpy medium:
the clump volume filling factor \( f_{c} \) and the ratio of the cloud to inter-cloud
medium density \( \alpha  \) (see V\'arosi and Dwek 1999\nocite{varosi99} for details).
In our simulations, we used a 40 kpc diameter spherical two-phase dusty medium
illuminated by a quasar central source with spectrum from the Cristiani and Vio \cite*{cristiani90}
quasar composite. The adopted total dust mass is \( 10^{8}M_{\odot } \) typical
of dust masses inferred from submillimeter observations (Best et al. 1998a\nocite{best98a},
Cimatti et al. 1998a\nocite{cimatti98a}, Archibald et al. 2000\nocite{archibald2000}) and
we used the standard Galactic dust model by Mathis, Rumpl and Nordsiek \cite*{mathis77}.
The result is displayed in fig. \ref{vary_tau} where the HzRG composite
spectrum (grey line) is compared to the scattered quasar (black continuous line).
This provides a remarkably good fit to both continuum and broad lines. We also
plotted the original quasar composite (dotted black line) as a reference to
illustrate the approximate greyness of the scattering. 

The best results were obtained for models with high clump to inter-clump medium
density contrast (\( \alpha =1000 \)) and low clump volume filling factor (\( f_{c}=0.01 \)).
The reason for this is that in order to have grey scattering, high density
clumps in an extended low density inter-clump medium are required. In this regime
radiative transfer is dominated by the dense clumps. Additionally, this ensures
that the direct view to the quasar is usually relatively unobscured, consistent with
the scarcity of highly reddened radio loud quasars (see eg, Simpson and Rawlings 2000\nocite{simpson2000}).
However it is interesting to note that this scenario also predicts that there
must be cases where our line of sight to the quasar intercepts one of these
high density clumps, in which case the direct view to the quasar would be completely
blocked. Scattered light would then dominate the flux in the UV-optical, resulting
in a type 1 object masquerading as a radio galaxy.

The main features of the continuum shape --- the dip around 2200\( \, 
\)\AA~ and the rollover below 1500\( \,  \)\AA --- are well reproduced
by this model (see fig. \ref{vary_tau}). It is difficult, however, to
use these observations to reliably constrain the dust properties since
the scattered spectrum is sensitive only to the difference in
wavelength dependence of the scattering and absorption cross-sections.
In particular the possible detection of the 2200\( \,  \)\AA~ dust
feature is difficult to confirm because it overlaps with the UV Fe\textsc{ii}
blends often present in quasar spectra above 2300\( \,  \)\AA~ (the so-called 
small blue bump). In addition, the signal-to-noise ratio in this
region of the spectrum is rather low due to strong sky lines (shaded
areas on fig. \ref{polfig}). The strength of these Fe\textsc{ii}s blends in quasar
spectra varies significantly from object to object and its origin is
still poorly understood. It seems, nevertheless, difficult to ascribe
the observed feature to Fe\textsc{ii} blends alone since it would require this
blend to be unusually strong in the spectrum of the hosted quasar in all sources, as
strong as that observed in ultra-strong Fe\textsc{ii} emitters like quasar
2226$-$3905 (Graham et al. 1996\nocite{graham96}). In consequence, the feature we
observe around 2200\( \,  \)\AA~ is probably a combination of the 2200\(
\,  \)\AA~ dust feature and UV Fe\textsc{ii} blends above 2300\( \,  \)\AA. 

Although inhomogeneous radiative transfer models considering anisotropic central
sources are not yet available, we can check that the predicted scattered fluxes
are consistent with our observations. The model described above predicts
that the fraction of scattered light in a spherical medium is of the of order
of 7\% of the input flux from the central source between 1000 and 3000\( \,  \)\AA.
This value is consistent with the Witt and Gordon \cite*{witt99a} {}``clumpy shell{}''
model predictions (their \( \tau _{v}=0.75 \) model). Assuming an average half
opening angle of 45\degr~ for the obscuring torus, we then estimate that about
2\% of the central AGN flux is scattered by the ISM of a radio galaxy. The R
magnitudes of the objects in our sample range between 21 and 23 and the inferred
magnitude of the central source is therefore between 17 and 19, consistent with observed
bright quasar magnitudes at \( z\sim 2.5 \).

The spatially-integrated level of continuum polarization is determined
by two processes: the geometrical dilution --- which is the result of
averaging over the possible scattering angles within the ionization
cones --- and dilution by sources of unpolarized radiation. The
appropriate biconical geometry in the framework of the unified scheme
has been studied in detail by Manzini and di Serego Alighieri
\cite*{manzini96} assuming an optically thin homogeneous medium.
Currently, none of the available radiative transfer models for
inhomogeneous clumpy media have considered the fractional polarization
of the scattered radiation in a biconical geometry. However, a study of
optical depth and geometrical effects on dust scattering by Zubko and
Laor \cite*{zubko99} shows that the fractional polarization is almost
independent of optical depth due to the relatively small contribution of
multiple scattering to the net polarization. In consequence, we assume
that the calculations made for homogeneous optically thin medium by Manzini
and di Serego Alighieri \cite*{manzini96} provide a reasonable estimate
of the fractional polarization of the scattered radiation. Considering
only calculations made for standard MRN interstellar dust composition
and a torus half opening angle \( \Phi =45 \)\degr, the fractional
polarization at a given wavelength is determined by the observer's
viewing angle \( \theta \). These models predict that the continuum
polarization at 1500~\AA~ varies between 25\% for \( \theta =90 \)\degr
(when the torus is seen edge on) and about 10\% for \( \theta =50
\)\degr. This range in fractional polarization depends on the assumed torus 
opening angle. The value we use in this calculation is illustrative and is 
consistent with Barthel's (1989)\nocite{barthel89} statistical analysis.

\begin{figure}
{\par\centering \resizebox*{0.98\columnwidth}{!}{\includegraphics{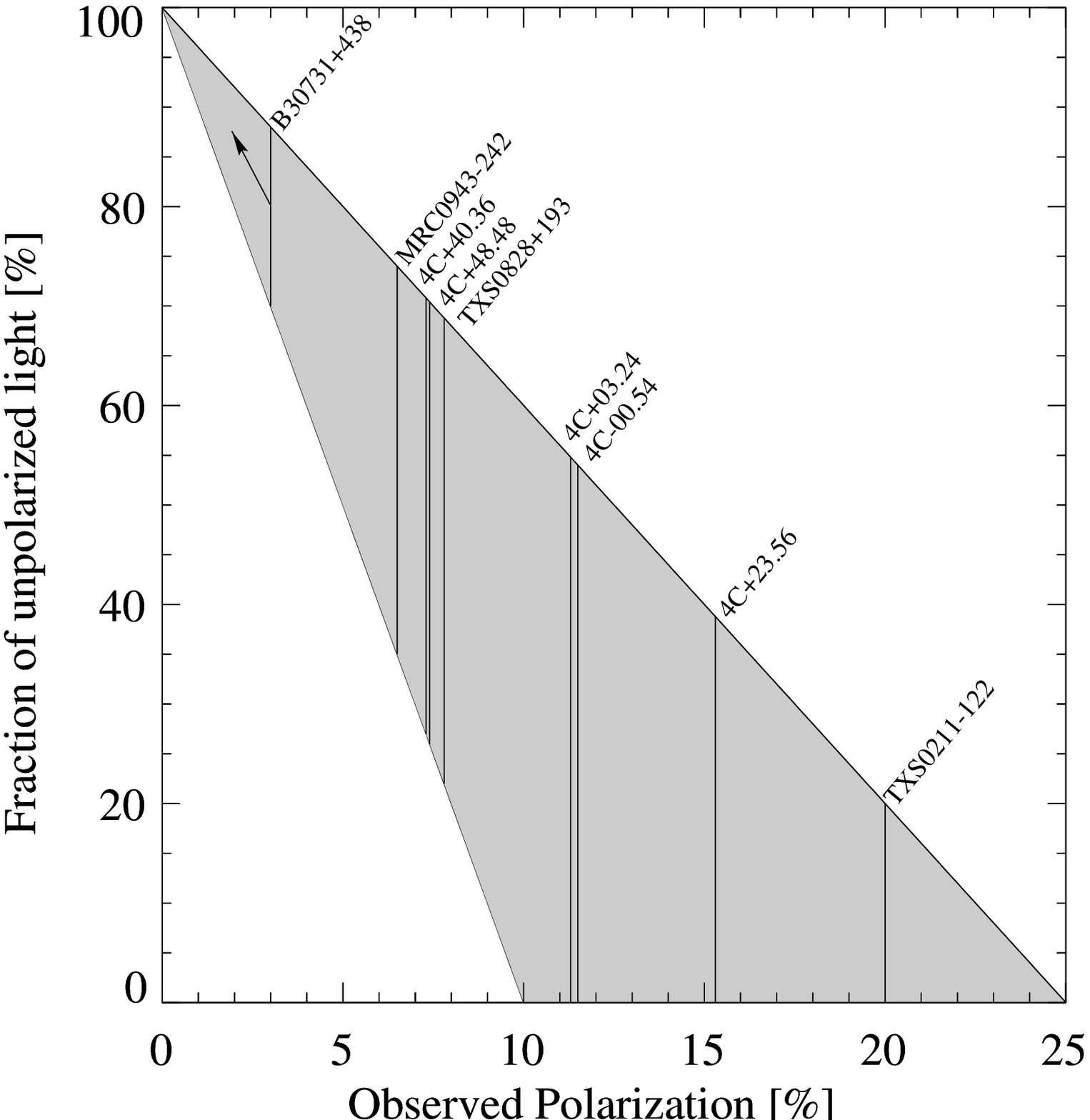}} \par}

\caption{\label{frac_unpol}Constraints on the fraction of unpolarized light contributing
to the continuum at \textasciitilde{}1500\protect\( \, \protect \)\AA. The
two lines delineating the shaded area represent \protect\( F_{unpol}\protect \) for extreme
values of the observer's viewing angle \emph{(top line}: \protect\( \theta =90\protect \)\degr;
\emph{bottom line}: \protect\( \theta =50\protect \)\degr). Vertical labeled
lines represent the possible range of variation of \protect\( F_{unpol}\protect \)
for each object in our sample. }
\end{figure}

The range in observed continuum polarization in our sample is compatible with
these predictions. It goes from less than 3\% in 0731$+$438 to about 20\% in 0211$-$122
longward Ly\( \alpha  \), which is lower than the range spanned by models. Using
the following equation describing the dilution of polarization of the scattered
radiation \( F_{scat} \) of fractional polarization \( P_{scat} \) by an unpolarized
component \( F_{unpol} \): \[
P_{obs}=P_{scat}\times \frac{F_{scat}}{F_{scat}+F_{unpol}}\]
we can set limits to the contribution of unpolarized radiation at \( \sim  \)1500\( \,  \)\AA~ making
extreme assumptions for the unknown observer's viewing angle as displayed on
fig. \ref{frac_unpol} (see also table \ref{model_parameters}). While the polarized
scattered radiation must contribute more than 80\% in 0211$-$122, unpolarized
radiation constitutes more than 70\% of the continuum in the least polarized
radio galaxy in this sample. All objects except 0731$+$438 are compatible with
scattered continuum accounting for at least half of the observed continuum.
The two main sources of unpolarized radiation --- the nebular continuum and a young
stellar population --- are discussed in the next two sections.

\subsubsection{The nebular continuum}

As shown by Dickson et al. \cite*{dickson95}, the nebular continuum
makes a significant contribution to the UV continuum of powerful radio
galaxies. Its spectrum, normalised to the observed \( H{\beta } \) flux, 
is computed using tabulated coefficients for \( T_{e}=10^{4}K \) given
in Aller et al. \cite*{aller87}. Since the \Lya~ flux is strongly
influenced by neutral hydrogen as well as dust, the \Lya/\( H{\beta }
\) ratio does not provide a reliable estimate of the \( H{\beta } \)
flux. We use instead the He\textsc{ii} line flux and the average value He\textsc{ii}\(
\slash  \)H\( {\beta }\sim 3.18 \) given by McCarthy \cite*{mccarthy93}
to compute the contribution of the nebular continuum. Assuming that both
the emission lines and the nebular continuum are affected by the same
amount of dust extinction, our estimate based on the He{\sc ii} flux
approximately takes this into account. The contribution to the total flux at 1500\( \, 
\)\AA~ for each object is given in the third column of table
\ref{model_parameters}. In general, the nebular continuum contributes
less than 25\% of the continuum at 1500~\AA~ leaving room for a
significant starlight contribution of unpolarized radiation.

\subsubsection{\label{ysp}The young stellar population}

\begin{table}
{\centering \begin{tabular}{cccc}
\hline 
Object&
Scattered &
Young stellar&
Nebular\\
&
AGN (\%)&
population (\%)&
continuum (\%)\\
\hline 
0211\( - \)122&
80-91&
11-0&
9\\
4C\( + \)23.56&
61-93&
32-0&
7\\
4C\( - \)00.54&
47-88&
41-0&
12\\
4C\( + \)03.24&
45-97&
52-0&
3\\
0828\( + \)193&
40-75&
35-0&
25\\
4C\( + \)48.48&
30-74&
58-14&
12\\
4C\( + \)40.36&
29-73&
61-17&
10\\
0943\( - \)242&
27-66&
53-14&
20\\
0731\( + \)438&
0-30&
84-54&
16\\
\end{tabular}\par}

\caption{\label{model_parameters}Range in the relative contribution of the different
continuum component at 1500\protect\( \, \protect \)\AA. }
\end{table}

\begin{figure*}
{\par\centering \subfigure{\resizebox*{0.48\textwidth}{!}{\rotatebox{90}{\includegraphics{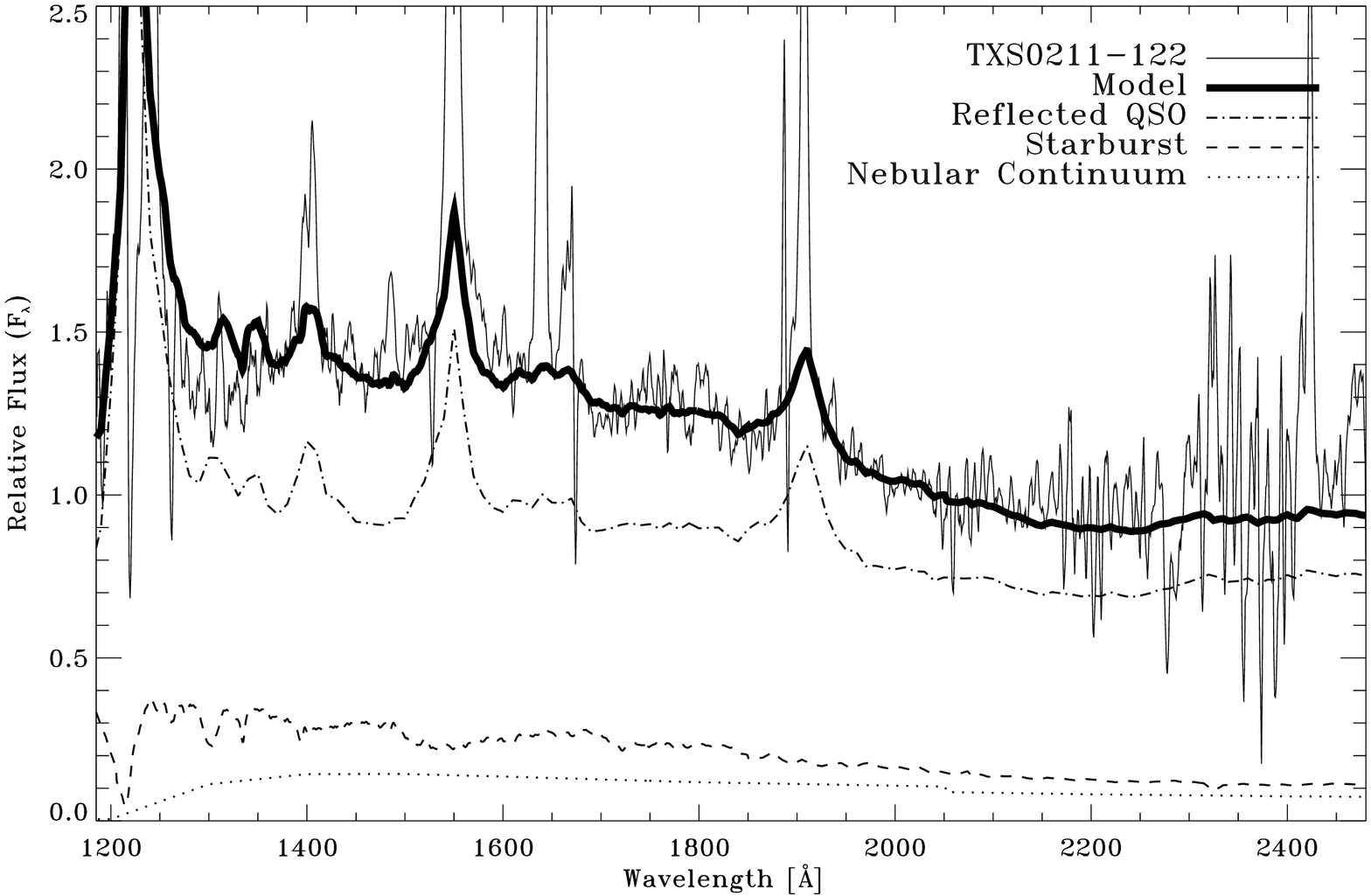}}}} 
\subfigure{\resizebox*{0.48\textwidth}{!}{\rotatebox{90}{\includegraphics{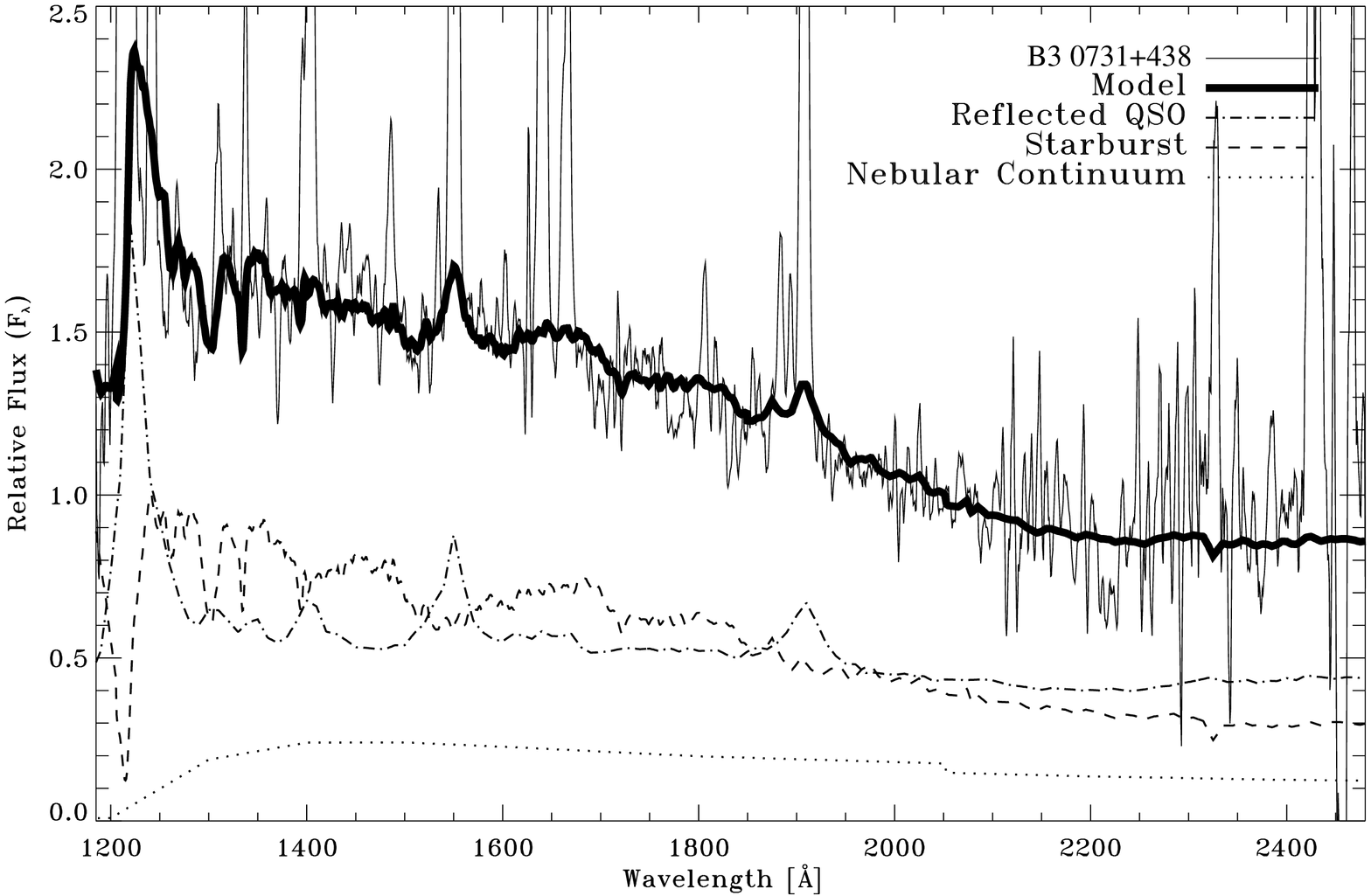}}}} \par}
\caption{\label{3comp_fit}Three components fit to the two objects in our sample showing
extreme polarization properties. \emph{Left}: 0211$-$122. \emph{Right}: 0731$+$438.}
\end{figure*}

From the constraints on scattered light and estimates of the nebular
continuum, we obtain limits on the fraction of continuum contributed
by a young stellar population. These limits are shown in the second
column of table \ref{model_parameters}.  In all objects except
0731$+$438, starlight contributes less than half of the UV
continuum. In consequence, it is not possible to accurately
characterise the underlying young stellar population.

Unambiguous direct spectral signatures of young
massive stars that we might expect to detect are
Si\textsc{iv}\( \, \lambda \, \)1400,
C\textsc{iv}\( \, \lambda \, \)1549 P~Cygni profiles and
purely photospheric absorption lines (ie. uncontaminated by
interstellar absorption) from O and B stars like Si\textsc{iii}\( \,
\lambda \, 1294 \), C\textsc{iii}\( \, \lambda \, 1427 \), Sv \( \,
\lambda \, 1502 \). We have searched for the presence of such features
in our sample, especially in low polarization objects where the
contribution of a young stellar population is likely to be the most
important.

We find no clear evidence for the presence of stellar P~Cygni profiles
in any of our spectra. These features are prominent in very young
stellar populations (\( t \lesssim 5 \) Myr) with equivalent width \(
W_{\lambda }\sim 8 \)~\AA~ (as measured on C\textsc{iv} line in
Leitherer et al. 1999\nocite{leitherer99} [Starburst 99] 1~Myr solar
metallicity Salpeter initial mass function (IMF) instantaneous starburst model). Such features
are prominent enough to be detected given the quality of our
data. However, the equivalent widths of these features vary strongly
with metallicity by a factor of about 5, the more metal-rich
starbursts having stronger lines (Heckman et al. 1998\nocite{heckman98}). 
In addition, the presence of strong narrow
emission lines and scattered broad lines could, in some objects,
partially fill such features if present. The blue wing of the
C\textsc{iv}\( \, \lambda \, \)1549 line profile is also affected by the
strong Si\textsc{ii}\( \, \lambda \lambda \, 1527,1534 \) absorption
and emission line.

Stellar population synthesis models and observations of
both local starbursts (eg. Leitherer et al. 1996 \nocite{leitherer96};
Conti et al. 1996\nocite{conti96}) and high redshift star forming
galaxies (eg. Pettini et al. 2000\nocite{pettini2000}) show that the
equivalent width of Si\textsc{iii}$\,\lambda\,1294$,
C\textsc{iii}$\,\lambda \,1427$ and S{\sc v}$\,\lambda\,1502$
photospheric absorption lines varies from 0.6 to 0.8\( \,  \)\AA. The
strength of these lines also depends strongly on metallicity. Heckman
et al. \cite*{heckman98} can only tentatively detect photospheric
features in their high signal-to-noise low metallicity composite
spectrum. We estimate the detection limit for an unresolved absorption
line to be about \( W_{\lambda }\sim 0.7 \)\( \,  \)\AA~ in the restframe, 
precisely in the expected range of observed values for stellar features. 
This means that in our observations we could only hope to detect these absorption
lines in the most favorable cases where young stars dominate the
continuum. Non-detections do not, therefore, provide useful
constraints. We do detect an absorption line that could be identified
as C\textsc{iii}$\,\lambda\,1427$ photospheric line in our spectrum
of 0731$+$438 (and also in 4C$-$00.54, see discussion in paper I). The
measured equivalent width in the restframe is 0.85\( \pm 0.2 \)\( \, 
\)\AA, consistent with values observed in starbursts. This purely
photospheric line is present only in the spectrum of early-B stars
\cite{mello2000} but since no other characteristic feature of young
massive stars is detected in any of our spectra, we consider 
this identification as very uncertain.

Constraints on the star formation rate (SFR) can be derived by comparing
the luminosity at 1500~\AA~ (listed in table \ref{results}), corrected
for scattered light and nebular continuum contributions (\( L^{\star
}_{1500} \)), to predictions from stellar population synthesis models.
We used a Starburst~99 (Leitherer et al. 1999)\nocite{leitherer99}, solar metallicity, Salpeter
IMF, continuous star formation model. In our
sample, \( log(L^{\star }_{1500} \)) ranges from about 42.2 to less than
40.7. Assuming that the stellar population is older than 10~Myr (ie. when
the stellar population has reached an equilibrium after the initial
onset of the burst), this translates into a range in SFR from
\textasciitilde{}\( 60\, M_{\odot }yr^{-1} \) to less than \( 2\,
M_{\odot }yr^{-1} \), not taking into account any dust reddening
correction. 

We could, in principle, combine the luminosity information with the
spectrum colour to constrain both the reddening and the age of the
stellar population. However, since reprocessed AGN radiation dominates
the continuum of all objects in this sample with the exception of
0731$+$438, the slope $\beta$ does not provide meaningful information
about the properties of the stellar population. In this one exception,
the slope at 1500~\AA~ \( \beta \sim -1.4 \), is significantly redder than
values spanned by starbursts models for which \( \beta  \) ranges from
-2.5 at 10~Myr to -2.0 at 1~Gyr. This shows that in this object the
stellar population is reddened with \( E_{B-V} \) values between
$\sim$0.1 and $\sim$0.2 if the age of
the stellar population (\( t^{\star } \)) is between 10~Myr and 1~Gyr. The
range in \( L^{\star }_{1500} \) for this object then translates into a
SFR between 30 and 120 \( M_{\odot }yr^{-1} \) for \( E_{B-V}=0.1 \) and
\( t^{\star }=10^{9}yr \) or between 60 and 260 \( M_{\odot }yr^{-1} \)
for \( E_{B-V}=0.2 \) and \( t^{\star }=10^{7}yr \) using a Galactic
reddening law (consistent with our scattering model) which gives a dust
extinction in magnitudes of \( A_{1500}=8.3\times E_{B-V} \).

This range in SFR appears to be comparable to what is typically observed
in Lyman break galaxies at a similar redshift. For instance  Sawicki and
Yee 1998\nocite{sawicki98} find that the median dust-corrected  SFR
in $z<3$ Lyman break galaxies is $\sim 560 M_{\odot }yr^{-1}$
(with a typical reddening $E_{B-V}\sim0.3$) or $\sim 20  M_{\odot
}yr^{-1}$ without dust correction. In the well studied lensed $z\sim2.7$
Lyman break galaxy MS~1512$-$cB58, Pettini et al.
(2000)\nocite{pettini2000} derive a SFR of about $70 M_{\odot }yr^{-1}$
(all values were converted to the cosmology used in this paper).

If we now compare to instantaneous burst models, we can set limits on
the total mass \( M^{\star } \) and the age \( t^{\star } \) of the
burst in 0731$+$438, depending on reddening. We obtain 
\( 4.9\,10^{9}<M^{\star }<2.2\, 10^{10}\, M_{\odot } \) and \( t^{\star }\sim
4.7\, 10^{7}yr \) without reddening, 
\( 2.6\, 10^{9}<M^{\star }<1.2\,10^{10}\, M_{\odot }
\) and \( t^{\star }\sim 1.6\, 10^{7}yr \) for \( E_{B-V}=0.1 \) and \(
5.6\, 10^{8}<M^{\star }<2.5\, 10^{9}\, M_{\odot } \) and \( t^{\star
}\sim 0.12\, 10^{7}yr \) for \( E_{B-V}=0.2 \).

As an example we present in fig. \ref{3comp_fit} a three component
fit (scattered quasar, nebular continuum and young stellar population)
of the two objects in our sample showing extreme polarization properties.
The contribution of each
component for these objects is given in table
\ref{model_parameters}. For 0211$-$122, --- the most polarized object ---
we display the fit that maximizes the stellar contribution allowed by
our model.  On the other hand for 0731$+$438 --- the least polarized
object --- we show the solution which maximizes the
scattered quasar's contribution. In both cases the stellar population was modeled
using the Starburst 99 solar metallicity 20~Myr instantaneous burst model.

The conclusion of this analysis is that the brightness of the scattered
quasar light in these powerful HzRG can swamp a SFR of several tens of
Solar masses per year unless a burst is observed during its initial,
very luminous phase. One at most of the objects in our sample appears
currently to be undergoing such a burst.

\subsection{Predicted far infrared continuum}

Current submillimeter instruments are just reaching the sensitivity
required to start measuring the far infrared (FIR) emission of brightest
HzRG and provide us with useful flux upper limits (Archibald et al.
2000\nocite{archibald2000}). As a consistency check, we have computed the FIR
SED that is expected from HzRG in a way that is consistent with our
dust scattering model. This estimate is restricted to the FIR flux
re-radiated by the same dust which is responsible for that scattering of
the ultraviolet continuum (ie. within the torus opening angle). 
Since much of this dust is located at large radial distances from the nucleus, 
it is relatively cool.

The characteristics of the ISM are the same as that used to model the scattered
continuum: a 40kpc sphere containing \( 10^{8}M_{\sun } \) of Galactic type
dust in an clumpy geometry with a density contrast \( \alpha =1000 \) and filling
factor \( f_{c}=0.01 \) is illuminated by an anisotropic central source. The
method is fully described in V\'arosi and Dwek \cite*{varosi99}.

This model does not provide a full treatment of the effect of the
anisoptropy of the central source nor does it include the effect of
radiation below the Lyman limit (\textasciitilde{}30\% total energy in
the UV). The latter would require modeling the ionization of gas and the
heating of dust by absorption of Ly\( \alpha  \) photons that is beyond
the scope of this paper.

Under these assumptions, the result of the simulations is the following energy
balance. The total input energy is \( \sim \, 10^{14}L_{\odot } \) --- the integrated
quasar ultraviolet-optical continuum above the Lyman limit. We are only interested
here in the fraction of this energy (29\%) that escapes within the torus opening
angle (the remaining 71\% is absorbed and re-radiated by the dusty torus at
wavelength around 10 \( \mu m \)). The \( \sim  \)\( 3\, 10^{13}L_{\odot } \)
going into the illuminated cone then divides into 77\% escaping directly without
interacting with dust grains, 7\% that is scattered (see sect. \ref{sect_scat}) and
16\% absorbed and re-radiated in the FIR. The resulting FIR luminosity of the
dust within the opening angle of the torus is about \( 5\, 10^{12}L_{\odot } \).
More than 75\% of this energy comes from the clumps and the rest from the interclump medium.

In order to compute the FIR SED, the temperature equilibrium for
graphite and silicate grains is computed by equating absorbed and
emitted luminosities in clumps and in the inter-clump medium using a
power law temperature probability distribution and assuming that that
the radiation density decreases with distance from the source like an
inverse powerlaw with index 2.5 (2 would be for optically thin case,
see V\'arosi and Dwek 1999).

\begin{figure*}
{\par\centering \resizebox*{0.98\textwidth}{!}{\rotatebox{90}{\includegraphics{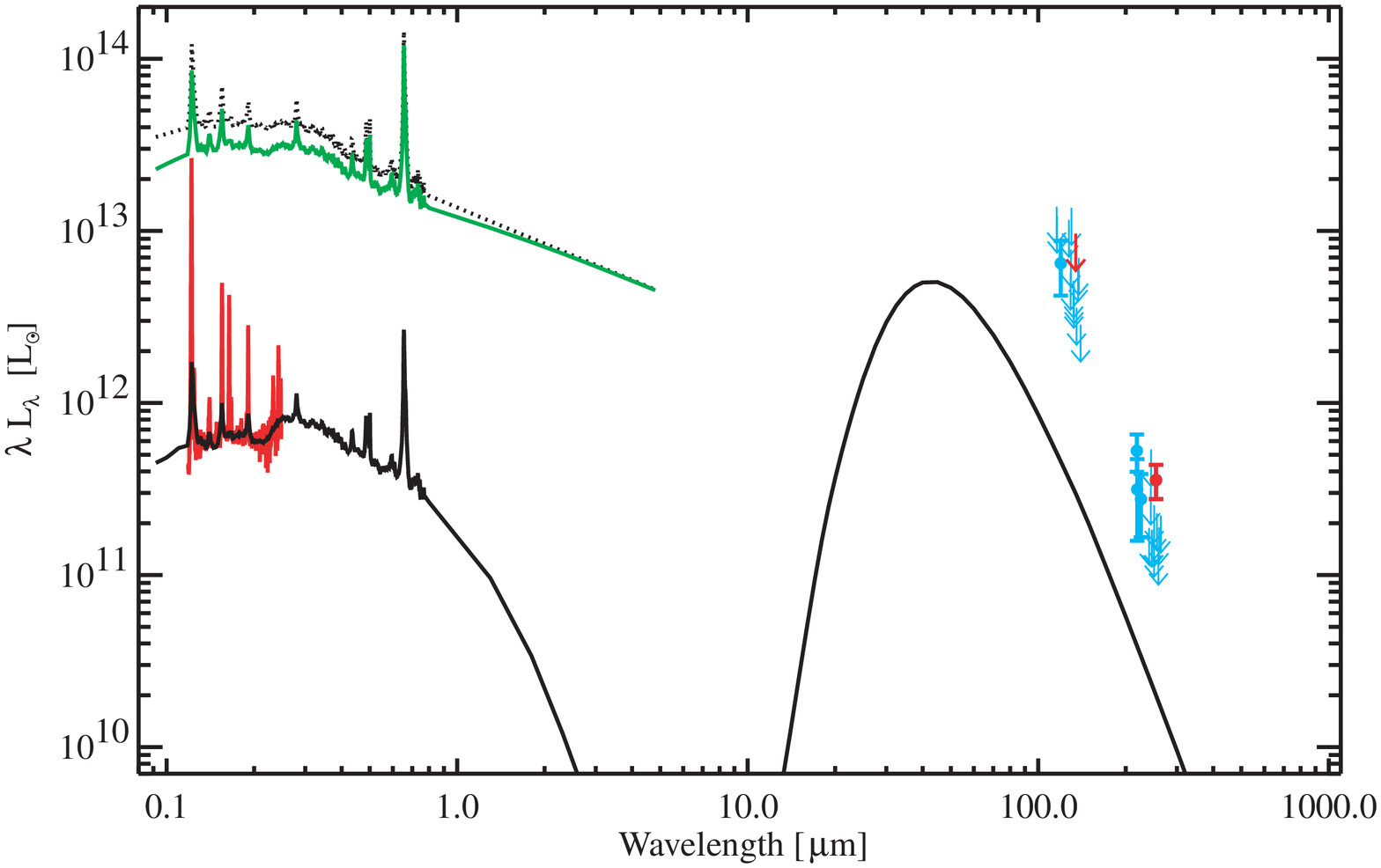}}} \par}

\caption{\label{fir sed}HzRG SED from 0.1 to 1000 \protect\( \mu m\protect \). \emph{Black
continuous line:} predicted scattered (\emph{left}) and FIR (\emph{right}) continuum;
\emph{black dotted line:} input UV-optical quasar spectrum; \emph{green line:}
radiation escaping directly; \emph{red line:} composite HzRG spectrum scaled
to 4C$+$48.48 luminosity. Dots and arrows show SCUBA measurements and 2\protect\( \sigma \protect \)
upper limits at 450 $\mu m$ and 850
\protect\( \mu m\protect \) (\emph{light blue}) of HzRG between redshift 2
and 3 from Archibald et al. 2000; \emph{red dot and arrow:} SCUBA measurements
for 4C$+$48.48 at 850\protect\( \mu m\protect \) and 450\protect\( \mu m\protect \).}
\end{figure*}
 
The resulting modeled SED from 0.1 to 1000 \( \mu m \) is plotted in fig.
\ref{fir sed}. The calculation shown here has been made to match the scattered
flux of 4C$+$48.48 from our Keck spectrum because this is the only object in our
sample detected with SCUBA at 850 \( \mu m. \) This corresponds to a quasar of
 magnitude \( R=17.3 \). The scattering is almost grey in the  UV but
at longer wavelengths, clumps become optically thin and the scattering deviates from grey and 
eventually the medium becomes transparent (the radiation that escapes directly
[green curve in fig. \ref{fir sed}] converges toward the input spectrum [dotted
curve] and the scattered radiation drops longward of 1\( \mu m \)). 

Our predictions are approximately a factor of ten below the upper limits.
It is important to stress at this point that what we compute here is clearly
a \emph{lower limit} to the FIR flux for several reasons. The contribution of
the dusty torus IR radiation is not included but it will mainly contribute around
10 \( \mu m \). In this calculation only the radiation of the dust \emph{within} the
cone opening angle is computed. The rest of the host galaxy (\textasciitilde{}70\%
of the ISM) mainly heated by star formation and reprocessed AGN radiation is
not taken into account although it clearly has a significant contribution at
long wavelength (\( \lambda >100\, \mu m \)). Also, as mentioned above, the transfer
of the radiation below the Lyman limit is not taken into account. 

The point of this calculation is not to obtain the best fit to the data but to show 
that our model, which is consistent with the observed scattered UV continuum, does 
not over-predict the FIR radiation.
This confirms the impression we have from observations that current instrumentation
is really at the sensitivity limit needed to detect HzRGs and that we currently
just pick the brightest objects. Most powerful radio galaxies 
at redshift between 2 and 3 should
be detected in the near future with an increase of instrumental 
sensitivity less than a factor of ten. 

\subsubsection{Emission lines}

The emission lines seen in radio galaxies arise from two quite distinct
regions. We see scattered light from the central quasar which is
characterised by its polarization and includes both broad and narrow
components. These narrow components arise in a region which is larger
than the broad line region (BLR) proper, but still small enough to be obscured from our
direct view, with the lines coming from transitions with a high critical
density for collisional de-excitation being more prominent (Hes, Barthel
\& Fosbury 1993\nocite{hes93}). The emission lines from the global ISM
of the host galaxy come predominantly from those zones which see the
ionizing radiation from the quasar directly and are generally much
stronger that the scattered narrow lines. While a number of mechanisms
may be responsible for heating and ionizing this gas, including shocks
and cosmic rays, the ionization of the warm ($\sim10^4$K) ISM in the
hosts of these powerful radio galaxies is thought to be dominated by the
AGN radiation field (Villar-Mart\'{\i}n et al. 1999\nocite{villar99}). The
presence of this galaxy-scale gas, illuminated by an intense ionizing
radiation field resulting in bright emission lines spanning a broad
range of ionization states, presents an opportunity for detailed studies
of its physical state and chemical composition which is absent in
galaxies without AGN. This opportunity cannot be easily exploited in the
quasars themselves since their spectra are so dominated by direct light
from the AGN. It is in the radio galaxies that we can most effectively
exploit the `natural coronograph' provided by the local obscuration of
the quasar.

Studies of chemical composition are likely to be particularly revealing since
this is the epoch where we expect the massive host galaxies and the AGN
themselves to be in the process of assembly or at least to be in an early
stage of their evolution. Since the timescales for AGN phenomena and galactic
evolution are very different --- by factors of a hundred or so --- we can
consider the quasar as acting like a `flash bulb' which gives us a snapshot
view of its host galaxy. By studying a number of such objects at a similar
redshift, we might expect these snapshots to reveal hosts in different stages
of evolution even if the galaxy and AGN formation triggers are closely
related.

In table \ref{em_lines}, we present measurements of all the detected emission lines. The
spectrograph aperture ranges from $4-8\arcsec\times1\arcsec$\ which corresponds
to 500--1000 kpc$^2$ at these redshifts. For objects of this size and
complexity, we must be realistic in our expectations for the results of the
analysis. The volume integrals represented by emission line measurements are
taken over a large fraction of the ISM of a whole galaxy and will include
inhomogeneities on many scales. The modelling we do vastly oversimplifies the
situation but nonetheless may allow us to interpret differences in line ratios
from object to object in terms of variations in their properties even if we
can place less confidence in, for example, absolute elemental abundance
determinations.

The spectra are characterised by a high state of ionization with the C{\sc
iv}, N{\sc v} and, in 0943$-$242 which has a sufficiently high redshift, O{\sc
vi} resonance lines being prominent. The He{\sc ii} recombination line is also
strong, comparable with C{\sc iv}. The signal-to-noise of the spectra is
sufficient to measure a number of intercombination lines from carbon,
nitrogen, oxygen and silicon. The average relative intensity of the strong
lines is similar to that in the average MRC radio galaxy spectrum compiled by
McCarthy (1993). The most significant variation amongst the stronger lines is
seen in \Lya, which is clearly affected by self-absorption, and in N{\sc v}.

\subsubsection{Photoionization modelling}

\begin{figure*}
{\par\centering 
\subfigure{\resizebox*{!}{9cm}{\rotatebox{0}{\includegraphics{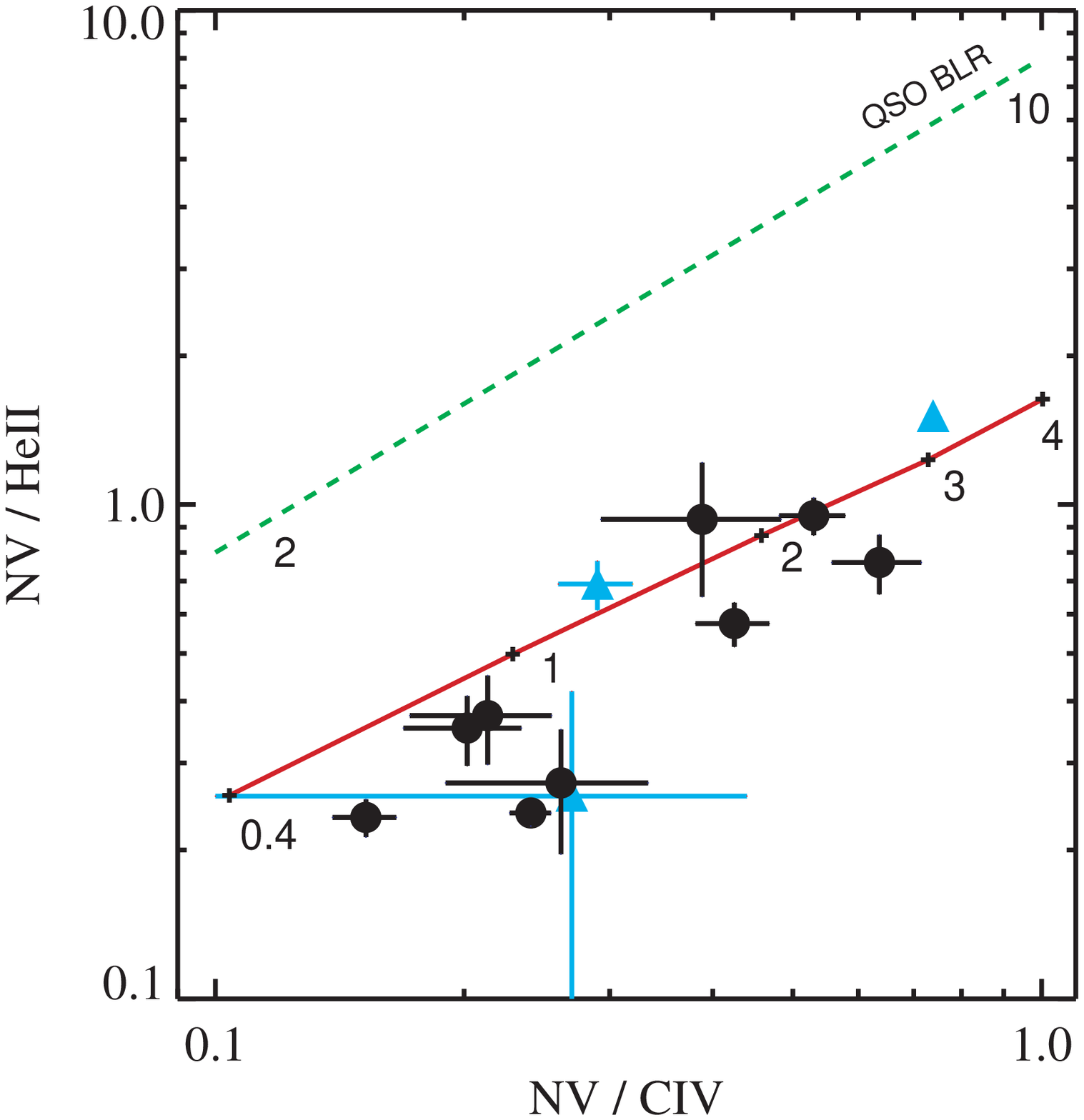}}}}
\caption{\label{nvciv_nvheii}The N{\sc v}/He{\sc ii} vs. N{\sc v}/C{\sc iv} diagram as plotted and
modelled by HF93 for the BELR of QSOs --- which clearly have much higher
densities than the gas we are sampling in the narrow lines. The dashed
line represents the locus of one of their chemical evolution models: M4a
which covers a range of 2--10 times solar for the primary elements in
this frame. The sequences implies solar or supersolar metallicities in
the extended gas of many HzRG and different levels of enrichment in
different objects. The three objects from the literature (see table
\ref{literature-results}) are plotted with different symbols
(triangles).}}
{\par\centering 
\subfigure{\resizebox*{!}{9cm}{\rotatebox{0}{\includegraphics{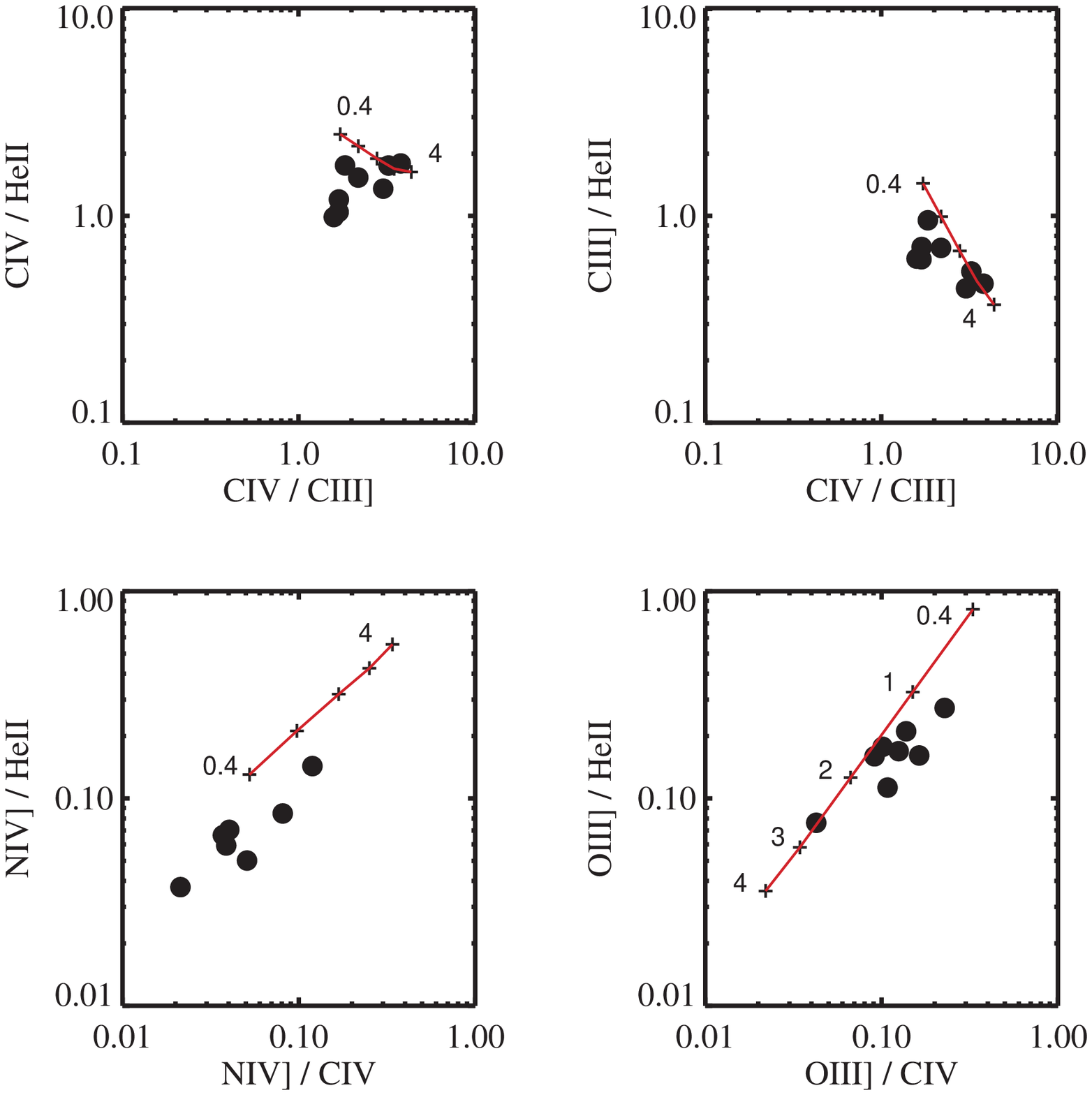}}}} \par
\caption{\label{diagnostic}
Diagnostic diagrams involving some important UV emission line
ratios including the intercombination lines 
N\textsc{iv}] $\lambda$ 1486, 
O\textsc{iii}] $\lambda\lambda$ 1661,1666 and 
C\textsc{iii}] $\lambda$ 1909. 
These include only data points from our sample. The model
metallicity sequence that best reproduces the data trend (solid line)
shows good agreement with the data in most diagnostic diagrams although
the N{\sc iv}] line predictions are substantially
displaced (lower left, see text). The numerical labels show the 
metallicity sequence from 0.4 to 4 times solar for elements other than
nitrogen which increases quadratically. The top two frames show the
small variation in ionization state within the sample while the oxygen
diagram (lower right) confirms the existence of the metallicity
sequence. In this and the associated figures, only emission line
components with a FWHM of $\le 3,000$ km s$^{-1}$ are included.
}}

\end{figure*}

A set of diagrams containing the line ratio data from our sample 
is shown in fig. \ref{nvciv_nvheii} and \ref{diagnostic}. The behaviour of the first of
these, the N{\sc v}/He{\sc ii} vs. N{\sc v}/C{\sc iv} diagram (fig. \ref{nvciv_nvheii}),
is reminiscent of the behaviour of luminous QSO broad emission line
regions discussed by Hamann and Ferland \cite*{hamann93,hamann99}
(HF93, HF99).  These authors showed that high redshift quasars ($z>2$)
define a tight correlation in this diagnostic diagram and the modeling
of the emission line ratios led them to conclude that the two N{\sc v}
ratios imply supersolar metallicities in the broad line region of many
high redshift quasars. They interpret the correlation in the N{\sc v}
diagram as a sequence in metallicity such that the highest redshift/most
luminous objects show the highest metallicities (up to 10$\times
Z_{\odot}$ or so). The reason for this could be related to the higher QSO and/or
host galaxy masses at large redshifts, implying a mass-metallicity
relationship in QSO similar to that seen in nearby ellipticals (Tinsley
1980\nocite{tinsley80}; Pagel \& Edmunds 1981\nocite{pagel81}; Vader
1986\nocite{vader86}; Bica 1988 \nocite{bica88}). This diagram is further
interepreted in terms of elliptical galaxy chemical evolution models by
Matteucci \& Padovani (1993).

The correlation defined by the sample of HzRG suggests a similar
interpretation: we might be witnessing different levels of metal
enrichment of the gas from object to object extending, possibly, to
supersolar metallicities.  If indeed we are seeing large metallicity
variations within our sample, the fact that it is apparent in the narrow
emission lines emitted by the galaxy ISM provides a much stronger
argument for global galactic chemical evolution than the small quantity of
nuclear gas represented by the quasar BLR.

In order to test the validity of this interpretation it was first
necessary to explore whether models other than a metallicity sequence
could reproduce the observations. Villar-Mart\'{\i}n et al. (1999)\nocite{villar99} 
studied the effects of shock ionization, AGN
photoionization and the influence of the AGN continuum shape, density
and/or ionization parameter $U$\footnote{$U$\ is the quotient of the
density of ionizing photons incident on the gas and the gas density:
$U=\int_{\nu_0}^{\infty}{\frac{f_{\nu} d\nu / h\nu}{c n_H}}$}.  They showed
that these models could explain neither the N{\sc v} correlation nor
the very strong N{\sc v} emission observed in some objects.  Other
effects investigated are the depletion of carbon onto dust grains and
the effect of illumination/viewing geometry on resonance line transfer
(see Villar-Mart\'\i n, Binette \& Fosbury 1996 \nocite{villar96}).
Carbon depletion can produce large N{\sc v}/C{\sc iv} ratios, but N{\sc
v}/He{\sc ii} remains nearly constant unless the metallicy is varied.
Regarding geometrical effects, since N{\sc v} and C{\sc iv} are resonance
lines, the contribution relative to other emission lines should vary
with the orientation of the object. However, N{\sc
v}/C{\sc iv} and N{\sc v}/He{\sc ii} vary in opposite ways with
perspective (N{\sc v}/C{\sc iv} increases when N{\sc v}/He{\sc ii}
decreases and viceversa) so that it is not possible to produce large
values ($\geq$1) of N{\sc v}/He{\sc ii} and N{\sc v}/C{\sc iv} at the
same time.

The effect on the N{\sc v} behaviour of a metallicity sequence was then
examined to see if it could explain the observations in this diagram
and remain consistent with the other emission line ratios. Thanks to the
high signal-to-noise of the spectra, we could use many fainter emission lines to to
compare with the models. It was assumed that the gas (100 cm$^{-3}$) is
photoionized by a power law of index $\alpha=-1.0$\ (Villar-Mart\'\i n et
al. 1999\nocite{villar99}) and the same $U (= 0.035)$\ was used for all the
objects. Such a constant ionization state is suggested by the small
variation of C{\sc iv}/C{\sc iii}] and C{\sc iv}/He{\sc ii}, fig.
\ref{diagnostic}).

It is found that:
\begin{itemize}
\item A sequence in metallicity can reproduce both the observed
correlation and the strength of the N{\sc v} emission.  The heavy
element abundances relative to H vary between 0.4 and 4$Z_{\odot}$ in the models.
\item  The N abundance must increase quadratically instead of linearly
with respect to carbon, oxygen etc.
\item  There is good agreement between the model  predictions  and the
data in all diagrams except for N{\sc iv}]$\lambda$1488.
This line is predicted to be stronger than observed. None of
the models we explored can explain this discrepancy. A similar
inconsistency was reported for the Seyfert galaxy NGC1068 (Kraemer
\& Crenshaw 2000\nocite{kraemer2000}). However, the fact that both the data
and the models define a tight correlation in the N{\sc iv}]$\lambda$1488 diagram (see
also O{\sc iii}] diagram) supports a metallicity sequence.
\end{itemize}

Another possibility, based on the polarization results, is
that the extended gas is mixed with dust. We studied the effects of
internal dust as a variation of the dust to gas ratio 
(depletion, scattering, absorption) and found that not
only are high metallicities needed, but also unrealistically high
densities (10$^6$ cm$^{-3}$) in the extended gas to explain the relative strength
of N{\sc v} in some objects. In addition, these models produce strong
discrepancies with other observed UV line ratios.  These results imply
that the emission lines come preferentially from regions where the
effects of internal dust are small.

We conclude that the N{\sc v} diagram implies solar or supersolar
metallicities in the extended gas of many HzRG. The sequence
shows that the level of enrichment varies from object to
object. The quadratic increase of N abundance suggests that
the N production is dominated by secondary nucleosynthesis processes, as
expected for high metallicities (Henry et al. 2000\nocite{henry2000}).

\subsubsection{Correlations}

Amongst the data for the continuum and the stronger emission lines, there are
two correlations which are immediately apparent. The first is between the
continuum fractional polarization $P$\ and the strength of
\Lya. Fig. \ref{lyaciv_p} shows that the highly polarized sources have
low
\Lya/C{\sc iv} ratios and fig. \ref{polfig}, especially the spectrum of
0211-122, suggests that the weakness of \Lya\ is due to self absorption as
has been demonstrated to occur in radio galaxies by van Ojik et
al. \cite*{ojik97b}.

{\begin{figure}

{\par\centering
\resizebox*{1\columnwidth}{!}{\rotatebox{0}{\includegraphics{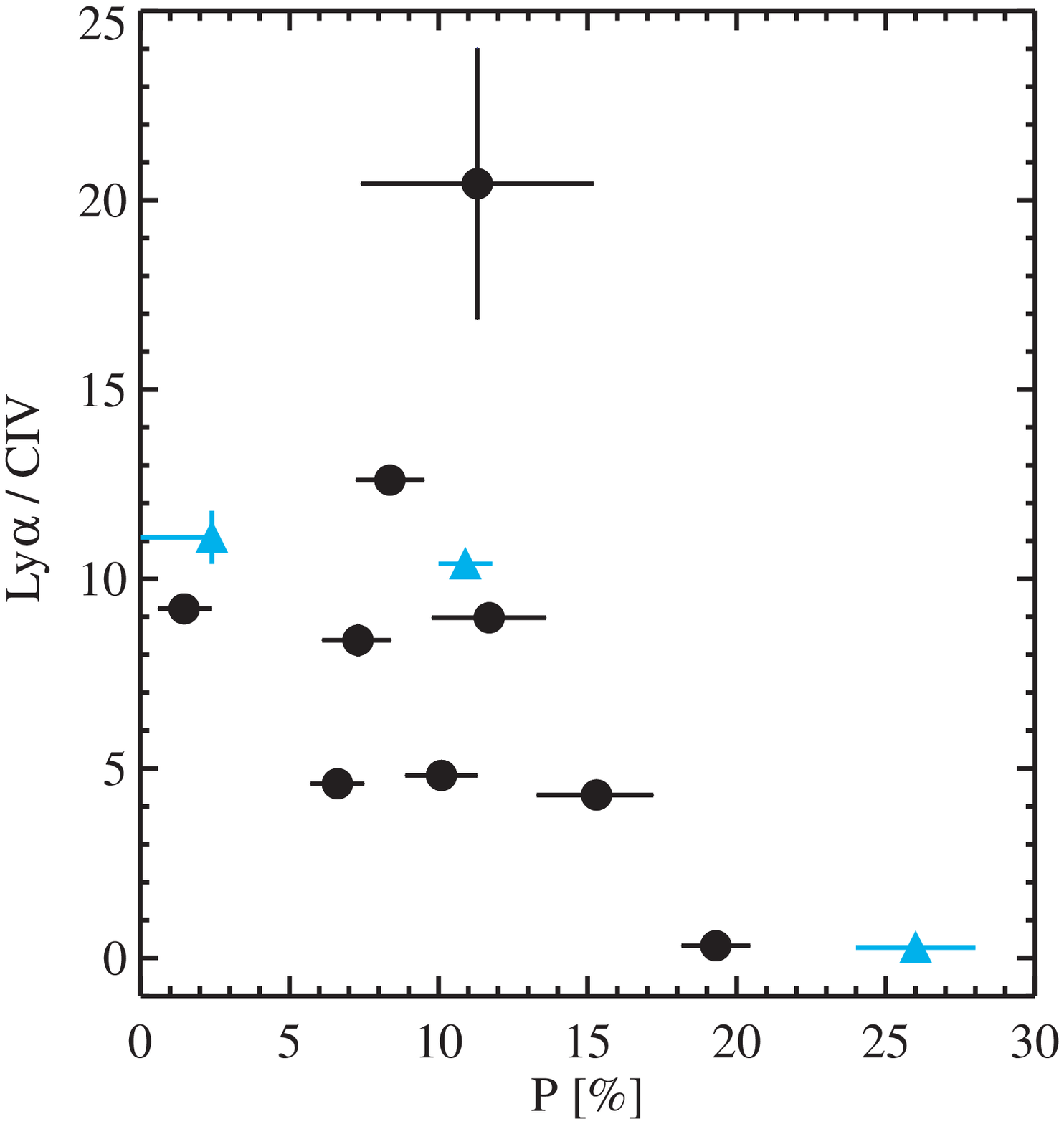}}} \par}
\caption{\label{lyaciv_p}The relationship between the strength of \Lya 
--- expressed as a ratio to C{\sc iv} --- and the continuum fractional
polarization measured in the bin just longward of \Lya. The error bars
represent 1$\sigma$ statistical uncertainties in $P$ but, for the line
ratios, are derived from uncertainties in the continuum fitting (see text).
For values of $P\lesssim3$\%, the points are plotted as $3\sigma$ upper
limits.}
\end{figure}
\par}
\vspace{0.3cm}

{\begin{figure}
{\par\centering
\resizebox*{1\columnwidth}{!}{\rotatebox{0}{\includegraphics{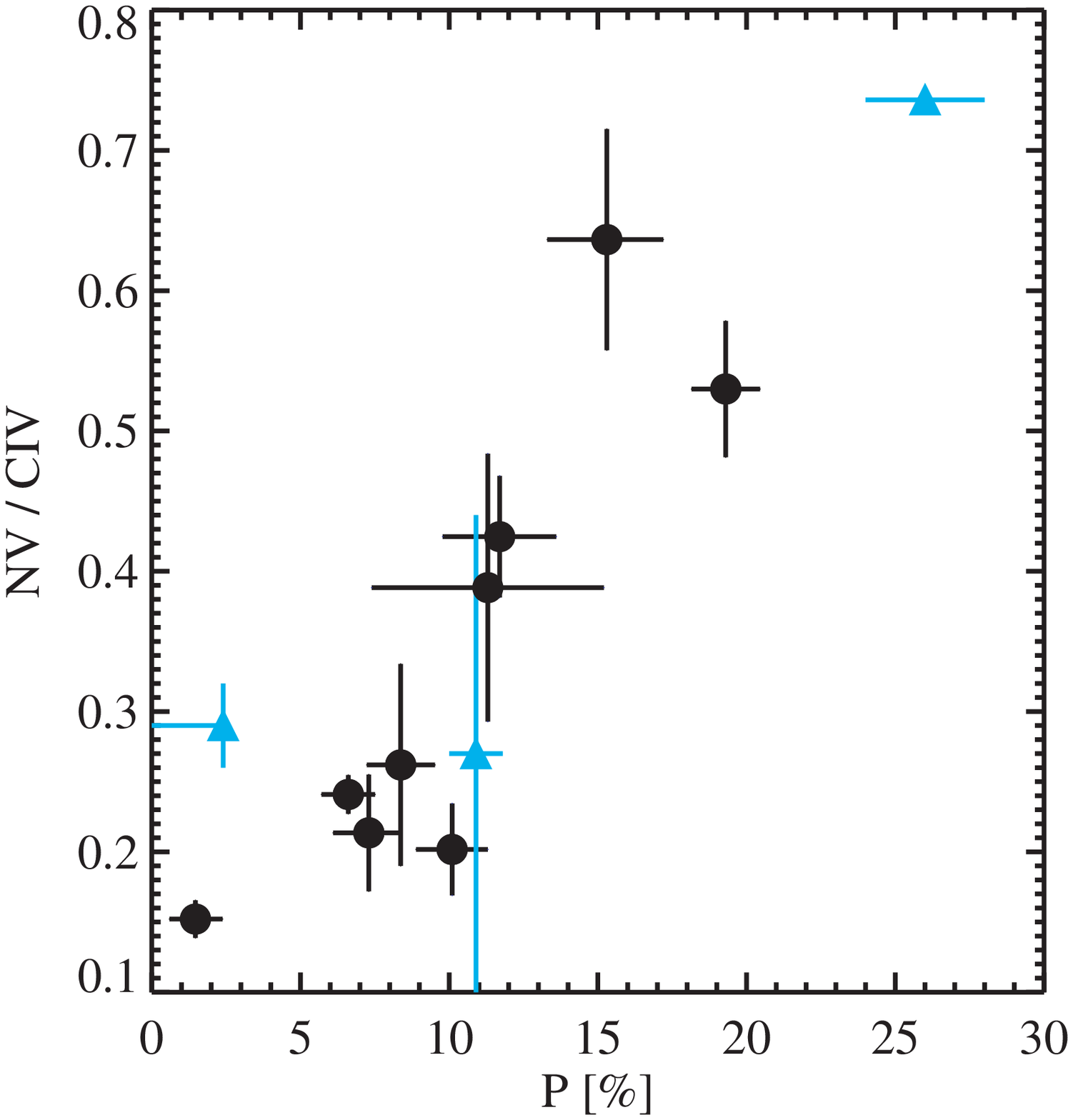}}} \par}

\caption{\label{nvciv_p}The relationship between the narrow line ratio
N{\sc v}/C{\sc iv} and the continuum fractional polarization measured in
the bin just longward of \Lya.}

\end{figure}
\par}
\vspace{0.3cm}

The second correlation, shown in fig. \ref{nvciv_p}, is between $P$\ and the
N{\sc v}/C{\sc iv} ratio with the latter spanning a factor of five within our
sample. The fact that this ratio is uncorrelated with C{\sc iii}]/C{\sc iv} or
He{\sc ii}/C{\sc iv} suggests that this is not simply an ionization
effect. The correlation is also present between N{\sc v}/He{\sc ii} and $P$,
which suggests --- as discussed in the previous section ---
that carbon depletion onto grains is not the dominant
effect. The tightness of this correlation is remarkable since, whatever the
underlying cause, these two physical quantities are likely to represent very
different averages over the emitting volumes. In the simplest case, the line
ratio is a microphysical property of the ionized gas while the continuum
polarization is dependent on large-scale structure and orientation.

The nature of these correlations can be further illustrated by comparing our
sample with other objects in the literature which have comparable quality
spectroscopy (see table \ref{literature-results}). Our most polarized source,
0211\( - \)122, has a line and continuum spectrum which is remarkably
similar to the lensed ultraluminous IRAS galaxy FSC~10214\( + \)4724
($z=2.282$, Goodrich et al. 1996\nocite{goodrich96}). Both sources have weak
(absorbed) \Lya, strong N{\sc v} and a high continuum polarization: 20\% for
0211\( - \)122 and 26\% for FSC~10214\( +
\)4724 measured in the same restframe band between N{\sc v} and Si{\sc iv}. 
Although FSC~10214$+$4724 is not a powerful radio source, the
Keck spectropolarimetry shows it to contain a luminous hidden quasar.

Our lowest polarization source, 0731\( + \)438, has strong \Lya\ and weak
N{\sc v} similar to 4C\( + \)41.17 ($z=3.798$) which Dey et al. \cite*{dey97a}
have argued has a continuum dominated by a young stellar population. The
relatively strong, low ionization interstellar absorption lines seen in the 4C
object appear, however, in emission in 0731\( + \)438. 

The $z=1.824$ radio galaxy 3C~256 has both imaging- (Jannuzi et al. 1995
\nocite{jannuzi95}) and spectropolarimetry (Dey et al. 1996\nocite{dey96}) with
$P$\ ranging from 11--18\% depending on aperture and wavelength. Palomar 5m
spectroscopy by Simpson et al. \cite*{simpson99} provides measurements of the
\Lya, N{\sc v}, He{\sc ii} and C{\sc iv} emission line fluxes. This object has
a continuum shape between 1500--2600~\AA~ which is essentially identical to our
average spectrum and is well fitted by the reflected quasar model.

These three sources have been plotted with symbols different from the main
sample in the figures.

In order to understand the origin of these correlations we need to
review the possible physical reasons for the variations of $P$, \Lya\
and N{\sc v}. For the scattering of quasar light by dust, the fractional
polarization we measure is determined by the particular biconical
scattering geometry and global orientation and by the presence of
diluting, unpolarized continua. The strength of \Lya\ with respect to
the other emission lines is probably most strongly influenced by neutral
hydrogen absorption with the possibility of dust playing a role in
quenching multiply scattered line photons (Villar-Mart\'{\i}n, Binette
and Fosbury 1996\nocite{villar96}). 

We have concluded that the
variation of the N{\sc v}/C{\sc iv} ratio is the result of abundance
variations. We cannot currently provide a
detailed explanation for this behaviour but it is likely, we believe, to
be a manifestation of an evolutionary sequence where a host galaxy
starts its assembly with low metallicity and being relatively dust-free.
As the spheroid building proceeds, the chemical enrichment of the ISM is
accompanied by dust production and dispersal. The similarity between the
spectra of 0211$-$122, our most polarized HzRG, and several of the
AGN-powered ULIRG, eg. FSC~10214$+$4724, suggests that these
dust-enshrouded objects are the {\em end point}\ of a
Gyr of spheroid assembly.

\section{Conclusions}

By performing spectropolarimetry of the restframe ultraviolet emission
from a sample of $z\sim2.5$\ radio galaxies, we have been able to
isolate and identify the dominant contributors to their observed
spectra. These are:

\begin{itemize}

\item{Spatially-extended, dust-scattered light from the underlying
obscured quasar. Due to the supposed clumpiness of the scattering
medium, a natural `luminosity-weighting' process operates which means
that the scattered photons originate predominantly from regions where
$\tau\sim 1$. This results in approximately grey reflection of a few
percent of the quasar luminosity. Deviations from greyness around
2200~\AA~ are expected for standard Galactic dust and are seen in our
data if the quasars do not have abnormally strong FeII emission. Based
on the spectral shape and the fractional polarization, the scattered
light accounts for between $\sim$90\% and $\sim$30\% of the total
continuum. In the most polarized sources, we also measure the broad line
emission from the quasar with approximately the expected equivalent and
velocity widths.}

\item{A nebular continuum which, based on the observed strength of the
HeII $\lambda 1640$\ recombination line, is typically $\sim$10\% of the
total flux in the UV.}

\item{The remaining continuum is assumed to arise from a young stellar
population although we have only very weak direct observational evidence
that this is the case. Constraints from the observed continuum colour
severely restrict the age and reddening of such a population in the low
polarization objects.}

\item{Strong, spatially-extended narrow emission lines from the ISM of
the host galaxy. From photoionization modelling studies it is shown that
the dominant fraction of this line emission is excited by the ionizing
radiation from the quasar.}

\end{itemize}

The clumpy dust-scattering model we have employed shows that a few
percent of the hidden quasar light is rendered visible in the restframe
UV spectrum of the radio galaxies. For luminous AGN, this is generally
sufficient to dominate the observed continuum in this wavelength range.
Only young, luminous and relatively unreddened starbursts will reveal
themselves clearly against the scattered `white haze'. Given the large
uncertainty in the reddening of starbursts and the consequent broad
range in the allowed conversion of stellar UV flux to SFR, it is likely
that we will have to await more sensitive sub-mm measurements in order
to investigate the consistency between the building of the stellar
content of the host spheroid and the chemical evolution of the ISM as
inferred from the emission lines.

The scattering models allow us to calculate directly the FIR radiation
emitted from the same dust within the ionization cones which reflects
the AGN. This is very much a lower-bound to the expected FIR flux since
it takes account neither of the warm dust emission from the obscuring
torus nor the cool dust emission from the shadowed dust outside the
cones which could be heated by obscured star-formation as well as 
reflected quasar light. We show that the
contribution from the scattering dust is typically 10\% of the observed
upper limits (and one 850$\mu$m detection for 4C$+$48.48) on sub-mm
emission from radio galaxies.

With the exception of \Lya\ and N{\sc v}, which show substantial
relative strength variation within our sample, the strong emission line spectra are very
similar and indicate that the ISM is ionized predominantly by the quasar
radiation field with an ionization parameter $U\sim 0.04$. The presence
of absorption components suggests that the variations in integrated
\Lya\ emission strength are the result of self-absorption by extended
neutral hydrogen halos. The variation by more than a factor of 5 in our
sample of the N{\sc v}/C{\sc iv} and N{\sc v}/He{\sc ii} emission line
ratios cannot be readily explained by variations in ionization state or
by the effects of carbon depletion onto dust. However, galactic chemical
evolution models for elliptical galaxies, where the nitrogen enrichment
arises predominantly from secondary processes over a timescale of
$\sim$1~Gyr, offer a good fit to the observations of these lines
and are generally consistent with the measurements of other lines. This
scenario is similar to that inferred from the BLR measurements of QSO at
similar redshifts which imply a range of supersolar metallicities from a
few to perhaps twenty times solar. The sequence for the radio galaxy ISM
suggests, however, a range of metallicity extending from about half to
a few times solar. In contrast to the QSO BLR results, where there appear
to be no metal-poor luminous AGN, the much more extended gas in the ISM
of radio galaxies shows that massive black holes can exist in relatively
metal poor hosts.

The N{\sc v}/C{\sc iv} and N{\sc v}/He{\sc ii} line ratios are
positively correlated with the continuum polarization $P$ while
\Lya/C{\sc iv} is anticorrelated with $P$. We believe that this is most 
likely to be the result of an evolutionary sequence within which the spheroids 
become increasingly dust-enshrouded as the chemical enrichment of the ISM
proceeds on a timescate of around 1~Gyr. A property of this scenario is
that ULIRG like FSC~10214$+$4724 are the {\em end point}\ rather 
than the beginning of such a period.

Our relatively low resolution spectra are not very sensitive to the
presence of weak, narrow absorption lines and so we cannot place very
stringent constraints on the presence of lines arising in the
photospheres of OB stars. We do, however, see a number of low ionization
resonance and excited fine structure lines from carbon, oxygen and
silicon. These lines are typically much weaker in absorption than seen
in Lyman-break galaxies like MS1512-cB58 (Pettini et al. 2000
\nocite{pettini2000}) and, indeed, sometimes show redshifted emission
components (see Fig.3). There is a weak tendency for high
polarization objects to show these lines in emission and low
polarization objects to show them in absorption. 
This could indicate the
different locations of the background stellar and scattered continuum
sources.

While we demonstrate that scattered quasar light generally dominates
the UV continuum of these powerful radio galaxies, the presence of an
evolved stellar population would be expected to reveal itself at
wavelengths above the 4000~\AA~ break. To this end, we are 
currently observing a partially overlapping sample of HzRG with
$2.2\leq z\leq 2.6$ in the near infrared with the ESO VLT in order to sample their
restframe optical spectra. Measurements of the continuum should reveal
the presence of any evolved stellar population while the familiar
optical forbidden emission lines will improve our ability to perform
reliable abundance analyses of the ionized ISM.

\acknowledgement{
We thank Luc Binette for all of his help and support in the use of the MAPPINGS code,
Adolf Witt for discussions about dust and Francesca Matteucci for her advice about
galactic chemical enrichment models. The W.M.Keck Observatory is operated as a scientific partnership between the
California Institute  of Technology and the University of California; it  was
made possible by  the generous financial support of the W.M.Keck Foundation.
 RAEF is  affiliated to the Astrophysics Division, Space Science Department,
European Space Agency.
}

\landscape{
\begin{table}[p]
\centering{
\begin{tabular}{llllllllllll}\\
\hline

Line				&$\lambda_{vac}$ (\AA)&4C$+$03.24&0943$-$242	&0828$+$193	&4C$+$23.56a	&0731$+$438	&4C$-$00.54	&4C$+$48.48	&0211$-$122	&4C$+$40.36\\
\hline
O\sc{vi}				&1031.9,1037.6	&		&5.6		&		&		&		&		&		&		&	\\
C{\sc ii} , C{\sc ii}$^{\star}$		&1036.3,1037.0	&		&5.2		&		&		&		&		&		&		&	\\
He\sc{ii}				&1084.9		&		&2.2		&		&		&		&		&		&		&	\\
\Lya				&1215.7		&254.4 $\pm$7.3	&213.8$\pm$1.8	&875.2$\pm$11.2	&77.4 $\pm$1.2	&428.1$\pm$2.2	&261.6$\pm$3.4	&700.0$\pm$11.8	&9.0$\pm$0.3	&1140.5$\pm$38.5\\
N\sc{v}				&1238.8,1242.8	&4.8 $\pm$0.8	&11.2 $\pm$0.6	&36.7 $\pm$5.9	&11.5$\pm$1.3	&7.1$\pm$0.6	&12.4$\pm$1.2	&14.5$\pm$4.0	&14.9$\pm$1.3	&29.1$\pm$5.6\\
\Lya + N{\sc v} broad		&		&		&		&		&22.5$\pm$5	&		&		&		&22$\pm$6	&	\\
Si{\sc ii}$^{\star}$			&1264.7,1265.0	&		&2.0		&		&		&1.0		&1.0		&6.4		&		&	\\
Si\sc{ii}$^{\star}$			&1309.3		&		&1.8		&3.0		&		&1.9		&		&		&0.9		&9.6	\\
C{\sc ii} , C\sc{ii}$^{\star}$		&1334.5,1335.7	&4.6 $\pm$1.3	&2.4		&3.3		&		&2.2		&1.1		&3.9		&		&20.9	\\
O{\sc iv}$]$, Si\sc{iv}		&{\it 1398}	&6.1 $\pm$2.5	&8.9		&31.4		&1.8		&9.7		&10.4		&13.4		&3.2		&22.0	\\
Si{\sc iv} broad			&1393.8,1402.8	&		&		&		&8.1		&		&		&		&6.5		&	\\
N\sc{iv}$]$			&1486.5		&		&2.4 $\pm$0.3	&7.3 $\pm$0.8	&2.2$\pm$0.6	&1.8$\pm$0.3	&		&4.5$\pm$0.8	&1.0$\pm$0.2	&2.9$\pm$0.6\\
Si\sc{ii}				&1526.7		&		&		&		&		&		&		&		&		&	\\
Si\sc{ii}$^{\star}$			&1533.4		&		&1.2		&		&		&0.9		&		&		&		&	\\
C\sc{iv}				&1542.9,1548.2	&12.5 $\pm$2.2	&46.5 $\pm$1.1	&181.7$\pm$4.5	&18.0$\pm$1.5	&46.5$\pm$1.0	&29.1$\pm$2.0	&55.5$\pm$0.5	&28.2$\pm$1.0	&136.1$\pm$5.1\\
C{\sc iv} broad			&		&		&		&		&13.9$\pm$1.8	&		&		&		&9.9$\pm$1.0	&	\\
$[$ Ne\sc{v}$]$			&1575.1		&		&		&0.9		&		&		&		&		&		&	\\
$[$ Ne\sc{iv}$]$			&1602.0		&		&		&2.2		&		&		&		&		&		&	\\
He\sc{ii}				&1640.4		&5.2 $\pm$1.3	&47.1$\pm$1.3	&103.8$\pm$1.4	&15.0$\pm$1.0	&30.4$\pm$0.5	&21.5$\pm$0.5	&53.2$\pm$3.5	&15.7$\pm$0.3	&77.7$\pm$5.6\\
O\sc{iii}$]$			&1660.8,1666.1	&		&7.6		&18.4		&4.1		&6.4		&3.6		&6.0		&1.2		&12.4	\\
N\sc{iii}$]$			&{\it 1749}	&		&1.4		&6.1		&		&		&		&		&		&8.4	\\
Si\sc{ii}				&1808.0		&		&1.7		&5.9		&		&1.4		&		&		&		&	\\
Si\sc{ii}$^\star$			&1816.9,1817.5	&		&		&1.4		&		&		&		&		&		&10.1	\\
Si\sc{iii}$]$			&1882.5		&		&2.9		&6.2		&		&1.7		&		&		&		&	\\
Si\sc{iii}$]$			&1892.0		&		&2.5		&12.0		&		&1.0		&		&		&		&	\\
C\sc{iii}$]$			&1906.7,1908.7	&		&29.3 $\pm$3.1	&55.9$\pm$6.8	&10.6$\pm$1.0	&21.2$\pm$0.5	&9.6$\pm$0.7	&32.8$\pm$1.0	&7.4$\pm$0.5	&73.9$\pm$11.8\\
C{\sc iii}$]$ broad			&		&		&		&		&12.2		&		&		&		&7.1		&	\\
C\sc{ii}$]$			&{\it 2326}	&		&		&19.2		&8.0		&5.0		&3.6		&11.6		&		&43.3	\\
$[$Ne\sc{iv}$]$			&2421.8		&		&		&49.0		&13.6		&18.0		&22.4		&32.6		&4.8		&37.1	\\
$[$O\sc{ii}$]$			&2471.0		&		&		&		&		&		&		&		&		&20.8	\\
\hline

\end{tabular}
}

\caption{\label{em_lines} Emission line flux measurements. Fluxes and errors are in units of  $10^{-17} \, $ergs$\,$s$^{-1}\,$cm$^{-2}$. Excited fine structure lines are labeled with a $\star$.  
Vacuum wavelengths are given in the second column. For blends of more than two lines, the average wavelength is shown in italics. Quoted errors for strong lines were computed as described in section \ref{section_emlines}.  
The typical error for weaker lines is about 10\%}
\end{table}

}

\endlandscape



\landscape{

\begin{table}[p]
\centering{
\begin{tabular}{llllllllllll}\\
\hline
Line				&$\lambda_{vac}$ (\AA)&4C$+$03.24&0943$-$242	&0828$+$193	&4C$+$23.56a	&0731$+$438	&4C$-$00.54	&4C$+$48.48	&0211$-$122	&4C$+$40.36\\
\hline
Si\sc{ii}				&1260.4		&		&		&		&3.7		&		&3.0		&		&7.5		&1.7	\\
Si\sc{ii}$^{\star}$			&1264.7,1265.0	&		&$-$13.4	&		&		&$-$4.9		&$-$4.6		&$-$12.5	&		&	\\
O\sc{i}, O\sc{i}$^{\star}$, Si\sc{ii}	&{\it 1304}	&		&		&		&4.2		&3.9		&5.5		&2.1		&		&	\\
Si\sc{ii}$^{\star}$			&1309.3		&		&$-$10.8	&$-$6.8		&		&$-$8.6		&		&		&$-$5.7		&$-$9.6	\\
C{\sc ii} , C\sc{ii}$^{\star}$		&1334.5,1335.7	&$-$31		&$-$12.4	&$-$6.8		&6.7		&$-$9.9		&$-$6.1		&$-$6.9		&2.2		&$-$20.9\\
Si\sc{ii}				&1526.7		&		&		&5.3		&3.1		&		&		&4.7		&6.8		&10.2	\\
Si\sc{ii}$^{\star}$			&1533.4		&		&$-$7.8		&		&		&$-$4.2		&		&2.7		&1.0		&	\\
Si\sc{ii}				&1808.0		&		&$-$7.9		&$-$18.7	&		&$-$8.4		&		&		&		&	\\
Si\sc{ii}$^{\star}$			&1816.9,1817.5	&		&		&$-$4.5		&		&		&		&		&		&11.6	\\
\hline

\end{tabular}
}

\caption{\label{low_ion_lines}Observed frame equivalent width ($W_{\lambda}$) measurements of low ionization zero-volt or excited fine-structure (labeled with a $\star$)  lines in \AA. 
$W_{\lambda}$ is positive for absorption lines and negative for emission lines.Vacuum wavelengths are given in the second column. For blends of more than two lines, the average wavelength is shown in italics.The equivalent width measurement error is typically about 10 \%}
\end{table}

}

\endlandscape


\bibliographystyle{aabib99}
\bibliography{aamnem99,all}

\end{document}